\RequirePackage{fix-cm}
\documentclass[twocolumn]{svjour3}          
\smartqed  
\usepackage{latexsym}
\usepackage{graphicx}
\usepackage{amsmath}
\usepackage{xcolor}
\usepackage{xspace}
\usepackage{mathtools}
\usepackage{multirow}
\usepackage{booktabs} 
\usepackage{soul} 
\usepackage{enumitem}
\usepackage{caption}
\usepackage{setspace}
\usepackage{bm}
\usepackage{makecell}
\usepackage{amssymb}
\usepackage{hyperref}
\usepackage{placeins}
\usepackage{transparent} 
\usepackage[linesnumbered,ruled,vlined]{algorithm2e}

\newcommand{\oursys}{\texttt{TableMark}\xspace}
\newcommand{\hi}[1]{\vspace{.25em}\noindent \textbf{#1}}
\newcommand{\myhi}[1]{{\it #1}}
\newcommand{\remove}[1]{}
\begin{document}

\title{TableMark: A Multi-bit Watermark for Synthetic Tabular Data
}


\author{Yuyang Xia  \and Yaoqiang Xu \and Chen Qian \and Yang Li \and    Guoliang Li \and
        Jianhua Feng
}


\institute{Yuyang Xia \\
              Yaoqiang Xu  \\
              Chen Qian  \\
              Yang Li  \\
              Guoliang Li \\
              Jianhua Feng
}

\date{Received: date / Accepted: date}

\maketitle

\begin{abstract}

Watermarking has emerged as an effective solution for copyright protection of synthetic data. However, applying watermarking techniques to synthetic tabular data presents challenges, as tabular data can easily lose their watermarks through shuffling or deletion operations. The major challenge is to provide traceability for tracking multiple users of the watermarked tabular data while maintaining high data utility and robustness (resistance to attacks). To address this, we design a multi-bit watermarking scheme \oursys that encodes watermarks into synthetic tabular data, ensuring superior traceability and robustness while maintaining high utility. We formulate the watermark encoding process as a constrained optimization problem, allowing the data owner to effectively trade off robustness and utility. Additionally, we propose effective optimization mechanisms to solve this problem to enhance the data utility. Experimental results on four widely used real-world datasets show that \oursys effectively traces a large number of users, is resilient to attacks, and preserves high utility. Moreover, \oursys significantly outperforms state-of-the-art tabular watermarking schemes.

\keywords{Copyright Protection \and Watermark \and Synthetic Tabular Data}
\end{abstract}

\section{Introduction}

The data market (e.g., Ali Data Market~\cite{AliDataMarket} and AWS Data Exchange~\cite{AWSDataExchange}), where data buyers purchase data products from data owners, significantly enhances the value of big data~\cite{DataMarket1,DataMarket2,DataMarket3,DataMarket4,DataMarket5}.\! Synthetic tabular data, which are artificially generated tables that closely resemble distributions of original tables~\cite{SyntheticTable1}, offer substantial utility to data buyers while safeguarding data privacy~\cite{SyntheticTable2,SyntheticTable3,SyntheticTable4}. These data products are crucial and widely used in data-driven decision-making and statistical analysis~\cite{SynMeter}. However, synthetic tables can be easily copied at a low cost, making them vulnerable once released. For example, an urgent problem is that data buyers may resell their copies of synthetic tables. Therefore, it is crucial to implement strategies that allow data owners to track and protect their data. This involves identifying and addressing unauthorized use to ensure that the data is shared responsibly and securely.

Recently, digital watermarking has emerged as an effective solution for copyright protection of synthetic data, demonstrating success in texts~\cite{LLMTextWatermarkSurvey} and images~\cite{GaussianSharding}. However, applying watermarking techniques to synthetic tabular data poses challenges~\cite{TabularMark}, as these schemes rely on patterns within large sets of tokens or pixels. Tabular data, consisting of small, independent tuples, can easily lose their watermarks through shuffling or deletion operations.

Generally, there are three vital properties of watermarking schemes for tracing synthetic tabular data in the data market.

(1) Traceability: The watermarking scheme should have sufficient capacity to encode multi-bit watermarks that allow for tracing multiple data buyers and can be applied to various data types.

(2) Robustness: The watermarking scheme should be resistant to attacks, which means that it should not be easily compromised by perturbations, modifications, deletions, and insertions.

(3) Utility: The quality of the watermarked data should remain high, meaning that the watermarked data should closely resemble the original data in terms of distributions~\cite{SyntheticTableSurvey} and support various tasks such as classification and regression~\cite{TabSyn}, as well as statistical analysis characterized by complex aggregation queries~\cite{SyntheticTable1}.

Existing schemes for watermarking synthetic tabular data can be divided into three categories. (1) Cell-based schemes encode watermarks into specific patterns within cells by first synthesizing non-watermarked tables and then modifying certain cells within these tables to form watermark patterns. (2) Statistic-based schemes encode watermarks into table statistics in a similar way, by modifying non-watermarked synthetic tables. (3) Generative schemes~\cite{MUSE} encode watermarks into pseudo-random sequences and synthesize watermarked tables from these sequences. However, existing schemes have some limitations. Firstly, cell-based schemes often have limited utility because modifications to cells, especially categorical ones, can compromise tuple integrity. For example, to ensure robustness, TabularMark~\cite{TabularMark} has to inject large-scale noise into a moderate proportion of cells. Secondly, statistic-based schemes generally provide better utility by imposing constraints on table modifications when encoding watermarks. However, they may fall short in traceability and robustness due to inappropriate constraints. For example, FreqyWM~\cite{FreqWM} imposes restrictive constraints that significantly weaken the encoded watermark patterns. Thirdly, generative schemes are weak in traceability, because it is hard to accurately recover pseudo-random sequences when decoding watermarks. For example, TabWak~\cite{TabWak} encodes watermarks into Gaussian noise vectors and inputs these vectors into a diffusion model to synthesize watermarked tables, but the process of recovering these noise vectors from watermarked tables is error-prone.

Designing a watermarking scheme for synthetic tabular data involves three challenges. The first challenge is determining an appropriate watermark channel and patterns, and designing an effective watermark encoding rule to encode the bit string watermark into watermark patterns within the channel, which ensures multi-bit watermarking to support a large number of data buyers, while preserving high utility. The second challenge is formulating the watermark encoding process as a constrained optimization problem, where utility is the optimization goal to maximize and a certain level of robustness is the constraint to satisfy, or vise versa, so that the data owner can effectively trade off robustness and utility. The third challenge involves designing effective optimization mechanisms to solve this problem to create watermarked tables that satisfy the desired trade-off between robustness and utility.

To address the first challenge, we partition the table and choose the partition histogram as the watermark channel. Specifically, to support multi-bit watermarking, we design multiple watermark patterns and their corresponding encoding rule based on relationships between histogram bin counts. For utility, on the one hand, we preserve tuple integrity by first calculating a histogram from the watermark to be encoded and then synthesizing a watermarked table according to this histogram, which ensures that individual tuples are not manipulated during watermark encoding. On the other hand, we maintain high-quality tuple-frequency distributions by minimizing differences between histograms of the watermarked and original tables. To address the second challenge, we allow the data owner to quantitatively express his/her robustness requirement and provide theoretical guarantees for it, by generating specific constraints on the histogram of the watermarked table to control the robustness of watermark patterns encoded in this histogram, and formulating a constrained optimization problem accordingly. To address the third challenge, we propose effective mechanisms to solve this problem to synthesize a high-utility watermarked table that also satisfies the robustness requirement.

Putting it together, a novel multi-bit watermarking scheme for synthesizing multiple sets of watermarked tabular data, \oursys, is proposed. Our contributions are summarized as below.

\noindent (1) We propose a novel multi-bit watermarking scheme \oursys designed for synthetic tabular data, which ensures superior traceability and robustness while maintaining high utility. (see Section~\ref{sec:overview})

\noindent (2) We establish constraints to provide theoretical guarantees for the data owner's robustness requirement and formulate the constrained optimization problem to balance robustness and utility. (see Section~\ref{sec:constraint_generator})

\noindent (3) We design effective optimization mechanisms to solve the constrained optimization problem. (see Section~\ref{sec:optimizer})

\noindent (4) We extensively evaluate \oursys using four widely used real-world datasets. The evaluation results demonstrate that \oursys significantly outperforms state-of-the-art baselines. (see Section~\ref{sec:experiments})

\setlength{\tabcolsep}{2pt}
\begin{table}[!t]
\centering
\caption{Summary of commonly used notations.}
\label{tab:notation}
{\fontsize{9pt}{11.5pt}\selectfont
\begin{tabular}{|c|l|}
\hline
\textbf{Notations} & \textbf{Descriptions} \\ \hline
$M$ & Number of clusters \\
$N$ & Maximum number of supported data buyers \\
$L$ & Length of bit string watermark \\
$D_o/D_w/D_s$ & Original/Watermarked/Suspect table \\
\rule{0pt}{2.5ex}$\mathbf{h}=(h_i)_{i=1}^M$ & Original histogram of $D_o$ \\
\rule{0pt}{2.5ex}$\mathbf{x}=(x_i)_{i=1}^M$ & Watermarked histogram to synthesize $D_w$ \\
\rule{0pt}{2.5ex}$\mathbf{y}=(y_i)_{i=1}^M$ & Extracted watermarked histogram from $D_s$ \\[1pt]
\hline
\end{tabular}
}
\vspace{-1.5em}
\end{table}

\section{Preliminary}

\subsection{Problem Definition}

In this paper, we consider the data market scenario, where the data owner has an original table and sells watermarked synthetic tables to data buyers. Specifically, watermarked tables are supposed to have the same cardinality as the original table to ensure high utility for downstream tasks (e.g., counting queries). For each data buyer, the data owner synthesizes a watermarked table that contains a watermark uniquely associated with this buyer, and sells the table to this buyer. For future identification of dishonest buyers, the data owner stores the buyer-watermark mapping in a watermark database. When the data owner spots a suspect table at a later point, s/he decodes a watermark from this table and matches this watermark against the database to identify potential dishonest data buyers. Note that we consider only the single-table case where the original and watermarked tables are single tables, and leave the multi-table case where the data owner sells watermarked multi-table databases as future work.

\subsection{Threat Model} \label{sec:threat_model}

We assume that the attacker (1) has access to the watermarked table, (2) has no access to the original table, table synthesis models, the watermark database, or the secret key of the data owner (the secret key is introduced for robustness and will be detailed in Section~\ref{sec:overview:cluster_pair_selector}), (3) has full knowledge of \oursys and its hyperparameters, and (4) aims to resell the watermarked table and thus does not excessively degrade its utility for various tasks when removing watermarks, to cater to the diverse needs of potential buyers.

\begin{figure}[!t]
  \centering
  \includegraphics[width=\linewidth]{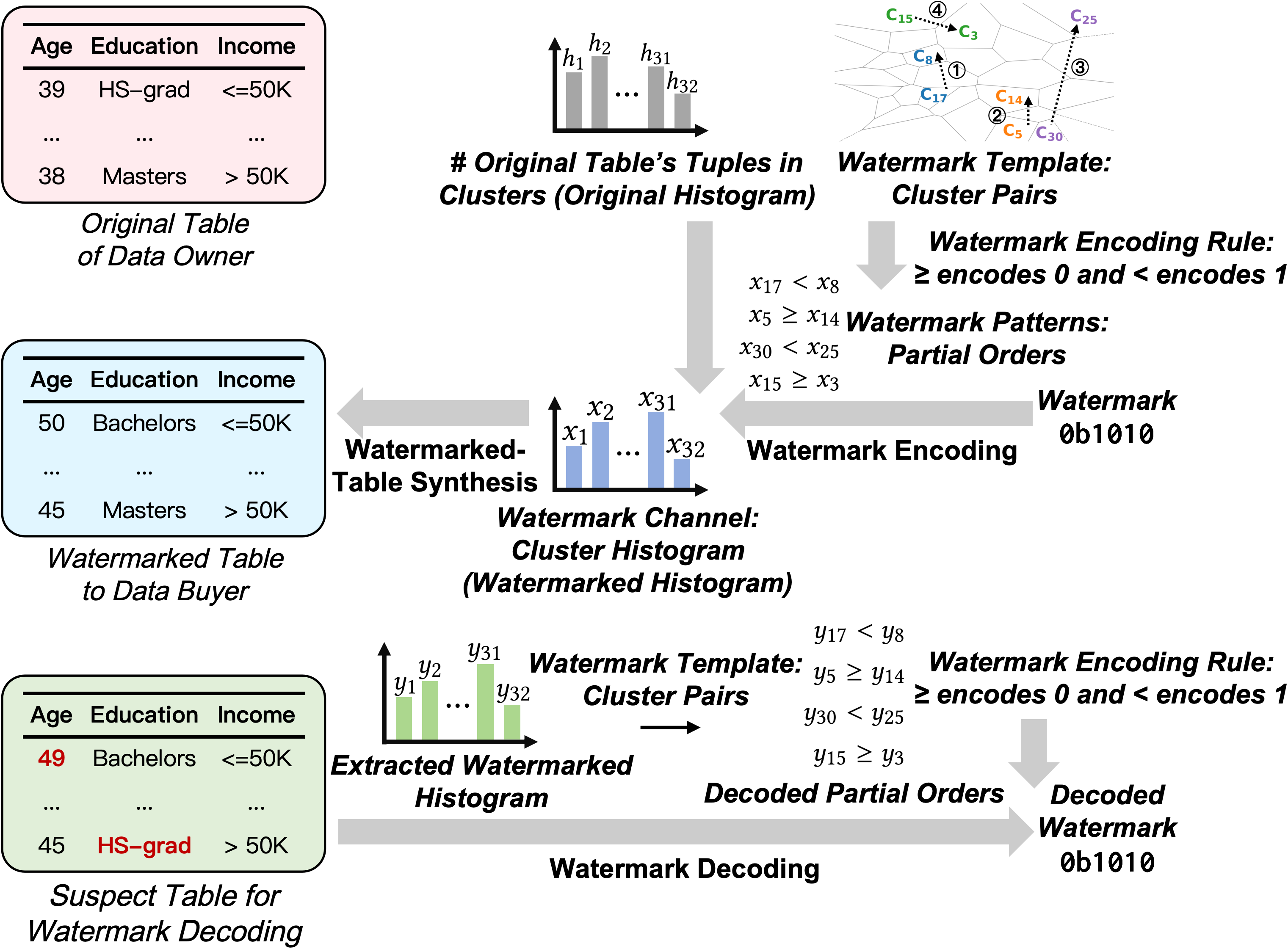}
  \caption{Terminologies of \oursys.} \label{fig:sketch}
\end{figure}

\begin{figure*}[!t]
  \centering
  \includegraphics[width=\linewidth]{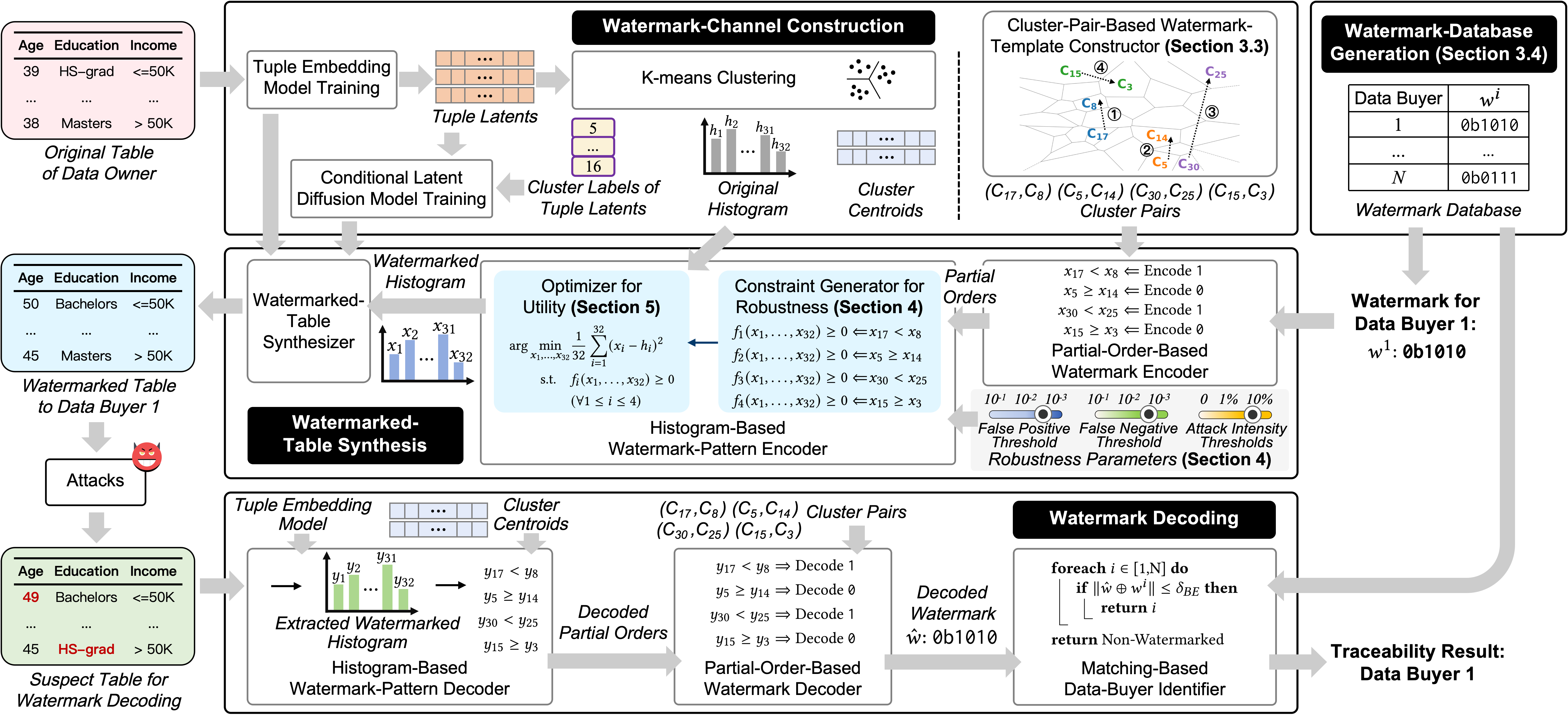} 
  \caption{Architecture of \oursys.}
  \label{fig:overview}
\end{figure*}

\section{Overview of \oursys} \label{sec:overview}

In this section, we provide an overview of \oursys. We first discuss the design principle in Section~\ref{subsec:watermark:overview} and then introduce the workflow of \oursys in Section~\ref{subsec:watermark:workflow}. Finally, we discuss important components of \oursys in Sections~\ref{sec:overview:cluster_pair_selector}-\!\!~\ref{sec:overview:histogram}. Table~\ref{tab:notation} presents a summary of commonly used notations. Figure~\ref{fig:sketch} illustrates terminologies and their relationships in \oursys.

\subsection{Watermark Overview} \label{subsec:watermark:overview}

\hi{Design Principle.} To balance traceability, robustness, and utility, we need to select appropriate {\it watermark channels} (i.e., where to encode the watermark) and {\it watermark encoding rules} (i.e., how to encode the watermark) in order to encode the {\it watermark} (i.e., what the watermark is) into the synthetic table. Specifically, the {\it watermark} is a fixed-length binary bit string, such as \texttt{0b1010}. The {\it watermark channel} refers to the table information that holds the watermark (e.g., individual tuples or table statistics). Given a watermark, the {\it watermark encoding rule} encodes specific {\it watermark patterns} (e.g., attribute correlations in individual tuples or relationships between table statistics) with respect to this watermark within the table information in the watermark channel. During watermark decoding, the watermark is decoded based on the patterns extracted from the suspect table $D_s$ and the encoding rule.

\hi{Watermark Channel.} Typically, individual tuples cannot serve as an effective watermark channel, because they have low entropy and cannot accommodate a large number of watermark bits. For traceability, we partition the original table $D_o$ and use multiple partitions as a foundation to construct the watermark channel. Specifically, we use clustering algorithms for table partitioning. This approach reduces the destructive impact of attacks on the watermark channel, because clustering algorithms tend to form well-separated clusters, so tuples are less likely to be perturbed across cluster boundaries. Considering that tuples in each cluster are unordered, we choose cluster statistics, specifically, cluster sizes (i.e., {\it cluster histogram}) as the {\it watermark channel} for robustness, because changes to any single tuple only lead to small fluctuations in the cluster histogram.

\hi{Watermark Template.} To encode $L$-bit watermarks, we construct $L$ {\it cluster pairs} from the clusters as the \myhi{watermark template} to align with the watermark. For example, in Figure~\ref{fig:sketch}, the watermark has four bits and the cluster pairs are $(C_{17},\!C_8),\!(C_5,\!C_{14}),\!(C_{30},\!C_{25}),\!(C_{15},\!C_3)$.

\hi{Watermark Pattern.} We conduct $L$ \myhi{partial orders} on the $L$ cluster pairs as {\it watermark patterns}, where each partial order is an inequality predicate ($\geq$ or $<$) between cluster sizes of a cluster pair, dictating a relative cluster-size relationship within the table to be synthesized that will hold the watermark. For example, in Figure~\ref{fig:sketch}, there are four partial orders: $x_{17}<x_8, x_5\geq x_{14}, x_{30}<x_{25}, x_{15}\geq x_3$, where $x_j$ is the size of cluster $C_j$ within the watermarked table. Note that partial orders among cluster sizes are more robust to cluster-size fluctuations than absolute cluster-size relationships.

\hi{Watermark Encoding Rule.} To encode $L$-bit watermarks into the watermark channel, we design a \myhi{watermark encoding rule} based on partial-order patterns: We use a single partial order to encode one bit ($\geq$ encodes 0 and $<$ encodes 1) and employ multiple partial orders to encode a bit string watermark. Thus, given an $L$-bit watermark, we generate $L$ partial orders on the $L$ cluster pairs as watermark patterns. Secondly, to encode the partial-order patterns into the watermark channel, we generate a cluster histogram (i.e., {\it watermarked histogram}) that will be used to synthesize the watermarked table $D_w$ to satisfy these partial orders. To achieve high utility, the watermarked histogram should closely resemble the {\it original histogram} of $D_o$. To achieve this goal, we generate the watermarked histogram by minimizing its difference from the original histogram, subject to the partial-order constraints.

\hi{Watermarked-Table Synthesis.}\; Using the watermarked histogram, we generate a watermarked table $D_w$\! whose cluster distribution aligns with this histogram.

\hi{Watermark Decoding.} Given a suspect table $D_s$, we predict the clusters for its tuples (called cluster assignment), calculate a cluster histogram (i.e., {\it extracted watermarked histogram}) from the cluster-assignment outcome, determine partial orders from the histogram and the cluster-pair template, and decode a watermark based on the partial orders and the watermark encoding rule.

\subsection{\oursys Workflow} \label{subsec:watermark:workflow}

\oursys involves four stages:\! \myhi{watermark-database generation}, \myhi{watermark-channel construction}, \myhi{watermarked-table synthesis}, and \myhi{watermark decoding}. Figure~$\ref{fig:overview}$ shows the architecture of \oursys. \texttt{Watermark-database} \texttt{generation} produces $N$ distinct $L$-bit watermarks to trace a maximum number of $N$ data buyers. \texttt{Watermark-} \texttt{channel} \texttt{construction} creates clusters for $D_o$, calculates the original cluster histogram, and constructs $L$ cluster pairs as the watermark template, which is shared among the watermark encoding and decoding processes for all buyers. \texttt{Watermarked-table} \texttt{synthesis} takes the watermark associated with the current data buyer, the $L$ cluster pairs, the original histogram, and the conditional generative model as input. It transforms the watermark into partial-order patterns according to the cluster pairs and the watermark encoding rule, and encodes these patterns into the watermark channel, by generating a watermarked histogram that resembles the original histogram while satisfying all partial-order constraints. Finally, it uses the conditional generative model to synthesize $D_w$ according to the watermarked histogram. \texttt{Watermark} \texttt{decoding} takes $D_s$, cluster pairs, and the watermark database as input, extracts partial orders from $D_s$, decodes a watermark from the extracted partial orders according to the cluster-pair template and the watermark encoding rule, and identifies the potential dishonest data buyer by matching the decoded watermark against the watermark database.

\hi{Watermark-Channel Construction.} To support heterogeneous data types, we embed tuples of $D_o$ into unified tuple latents (i.e., compact vector representations of tuples) using a variational autoencoder~\cite{VAE}, following~\cite{TabSyn}. To form multiple clusters, we perform K-means clustering on the tuple latents. The clustering results include the cluster label for each tuple latent, the original histogram, and cluster centroids. Then, we train a conditional generative model that is used to generate tuples for $D_w$ based on cluster labels. Specifically, we construct (tuple latent, cluster label) pairs from the clustering results as training data to train a conditional latent diffusion model, because latent diffusion models~\cite{LDM} exhibit superior performance in table synthesis tasks~\cite{TabSyn,SiloFuse}. Next, to obtain $L$ cluster pairs where each pair corresponds to a partial order that encodes a bit of the $L$-bit watermark, a naive method is to randomly select a subset of clusters and pair them into $L$ cluster pairs. However, for better traceability, robustness, and utility, we design an algorithm to construct high-quality cluster pairs, which will be described in Section~$\ref{sec:overview:cluster_pair_selector}$.

\hi{Watermark-Database Generation.} To trace a maximum number of $N$ buyers, we generate $N$ distinct $L$-bit watermarks and store them in the watermark database. The watermarks can be generated randomly, but we design an effective algorithm for better traceability and utility, which will be introduced in Section~$\ref{sec:overview:watermark_generator}$.

\hi{Watermarked-Table Synthesis.} Consider synthesizing $D_w$ for the $i$-th data buyer. We first assign a watermark $w^i$ to this buyer, which has not been previously assigned to other buyers, and record the buyer-watermark mapping. We then construct partial-order patterns based on the cluster-pair template, where each bit of $w^i$ is transformed into a partial order according to the watermark encoding rule. For example, as shown in Figure~$\ref{fig:overview}$, $L=4$ and the cluster pairs are $(C_{17},\!C_8)$, $(C_5,\!C_{14})$, $(C_{30},\!C_{25})$, $(C_{15},\!C_3)$. To synthesize $D_w$ for the first data buyer with the watermark \texttt{0b1010}, four partial orders are generated: $x_{17}\!<\!x_8, x_5\!\geq\! x_{14}, x_{30}\!<\!x_{25}, x_{15}\!\geq\! x_3$, where $x_j$ is the cluster size of $C_j$ within $D_w$ and $<$ $(\geq)$ corresponds to the watermark bit 1 (0).

To generate the watermarked histogram, a naive method is to solve a constrained optimization problem, where bin counts of the watermarked histogram are decision variables, a distance metric between the watermarked and original histograms (e.g., the mean squared error) is the objective to minimize, and partial orders are constraints. However, this approach has limitations in terms of robustness, because cluster sizes between partial orders may be similar in the watermarked histogram, making partial-order patterns sensitive to cluster-size fluctuations within $D_w$ due to attacks. To address this issue, we design effective algorithms for watermarked-histogram generation, which will be detailed in Section~$\ref{sec:overview:histogram}$.

To synthesize $D_w$ according to the watermarked histogram, for cluster $C_j$ with size $x_j$ in the watermarked histogram, the conditional latent diffusion model takes the label (in the form of one-hot embedding) of cluster $C_j$ as input and outputs $x_j$ tuple latents that fall in this cluster. These tuple latents are then reconstructed into tuples by the tuple embedding model to produce $D_w$. Note that the generation process of individual tuples is not manipulated by watermark encoding, which ensures tuple integrity and contributes to the high utility of watermarked tables.

\noindent \textbf{Watermark Decoding.} To identify the potential dishonest data buyer from $D_s$, we first embed tuples of $D_s$ into latents using the tuple embedding model, and assign tuple latents to clusters based on cluster centroids obtained during watermark-channel construction, following the K-means clustering paradigm. Secondly, we calculate the extracted watermarked histogram from the cluster-assignment outcome. Thirdly, we compare cluster sizes between the $L$ cluster pairs generated during watermark-channel construction to extract $L$ partial orders. Fourthly, we decode a watermark from the extracted partial orders, where a bit is decoded from each partial order according to the watermark encoding rule, and $L$ bits form the decoded watermark $\hat{w}$. For example, as shown in Figure~\ref{fig:overview}, the extracted partial orders are $y_{17}<y_8, y_5\geq y_{14}, y_{30}<y_{25}, y_{15}\geq y_3$, with $y_j$ denoting the size of cluster $C_j$ in the extracted watermarked histogram. Then $\hat{w}$ is \texttt{0b1010}. Finally, we match $\hat{w}$ with watermarks $w^i\;(1\leq i\leq N)$ in the database. If there exists $w^i$, such that the Hamming distance between $\hat{w}$ and $w^i$ does not exceed the bit error threshold $\delta_{BE}$, then $D_s$ is concluded to have been sold to the $i$-th data buyer. Otherwise, $D_s$ is concluded to be non-watermarked. Note that $\delta_{BE}$ is introduced to enhance robustness and is determined based on the false positive rate, which will be detailed in Section~$\ref{sec:constraint:trace}$.

\subsection{Cluster-Pair-Based Watermark-Template Constructor} \label{sec:overview:cluster_pair_selector}

Three considerations are important when constructing the cluster-pair template. For traceability, we need to maximize the information per watermark bit to trace a large number of data buyers in a few bits. Hence, we need to maximize the independence between watermark bits, which means that the decoding correctness of an individual bit is weakly correlated with that of other bits. Therefore, the partial orders used to encode different bits should be supported by distinct clusters, necessitating a total of $2L$ clusters to form the $L$ cluster pairs. For robustness, the pairing relationships of the selected clusters must be hidden from the attacker so that s/he cannot remove the watermark by manipulating tuples of specific cluster pairs. Thus, we need to pair the selected $2L$ clusters in a pseudo-random manner using the data owner's secret key as the seed, which can be implemented by shuffling ordinals of the $2L$ clusters and grouping these ordinals into $L$ non-overlapping pairs. For utility, the watermarked and original histograms should be close. However, since the partial orders in $D_w$ and $D_o$ may differ (called partial-order conflicts), forcing the watermarked histogram to deviate from the original histogram, it is desired that $D_o$ has similar numbers of tuples in the left-hand and right-hand clusters of the cluster-pair template. These cluster-size similarities with regard to $D_o$ ensure that partial-order conflicts result in only minor deviations in the watermarked histogram compared to the original histogram.
However, due to the pseudo-random pairing strategy, any two clusters from the $2L$ clusters may form a cluster pair, so we need to select $2L$ clusters with the minimal size variance with respect to $D_o$, formulated as below.

\begin{equation}
\begin{aligned}
\arg\min_{\substack{1\leq i_1<\dots<i_{2L}\leq M}} \operatorname{Var}(h_{i_1}, h_{i_2}, \ldots, h_{i_{2L}})
\end{aligned}
\end{equation}
where $h_i$ is the size of cluster $C_i$ within $D_o$.

The above problem can be solved using a sliding-window algorithm in $O(M\log M)$ complexity, where we first sort the array $h_1, \ldots, h_M$ and then slide a window of size $2L$ over the sorted array. We then compute the cluster-size variance for each window and finally select the clusters in the window with the minimal variance.

\subsection{Watermark-Database Generator} \label{sec:overview:watermark_generator}

To generate watermarks for tracing data buyers, two considerations are important. For traceability, any decoded watermark $\hat{w}$ should match at most one generated watermark given the bit error threshold $\delta_{BE}$ to ensure unambiguous tracing. Thus, the Hamming distance between two watermarks should be at least $2\delta_{BE}+1$. For utility, since different watermarks are transformed into different sets of partial orders that correspond to different numbers of partial-order conflicts, the generated watermarks should cause as few partial-order conflicts as possible. Furthermore, to ensure utility for every data buyer, the maximum number of partial-order conflicts should be minimized. Note that the number of partial-order conflicts caused by the watermark $w^i$ is exactly the Hamming distance between $w^i$ and the optimal bit string watermark $w^*$ that does not cause partial-order conflicts, where $w^*$ is calculated from the original histogram of $D_o$ and the cluster-pair template. For example, suppose that $L=4$, bin counts of the original histogram are $h_i\;(1\leq i\leq 32)$, and the cluster pairs are $(C_{17},C_8)$, $(C_5,C_{14})$, $(C_{30},C_{25})$, $(C_{15},C_3)$, then the four bits of $w^*$ are the truth values of the four partial orders: $h_{17}<h_8,h_5<h_{14},h_{30}<h_{25},h_{15}<h_3$, which align with the watermark encoding rule (i.e., $\geq$ encodes 0 and $<$ encodes 1) and thus do not cause partial-order conflicts. Therefore, the watermark generation problem can be formulated as follows.

\begin{equation}
\begin{aligned}
\arg\min_{w^1,\dots,w^N}&\max_{i\in [1,N]} \|w^i\oplus w^*\|\\
\text{s.t.}\quad &\|w^i\!\oplus\! w^j\|\!\geq\!2\delta_{BE}\!+\!1&&\!(\forall \; 1\!\leq\! i\!<\!j\!\leq\! N)\!\!\!\!\!
\end{aligned}
\end{equation}
where $\|w^i\oplus w^j\|$ denotes the Hamming distance between $w^i$ and $w^j$.

Note that the above problem involves $w^*$, which varies with each dataset (since different datasets correspond to different original histograms), making it data dependent. In other words, we need to solve this complicated problem separately for every dataset. To address this issue, we transform it into a data-independent problem to apply its solution to any specific dataset. Let $\bar{w}^i\!=\!w^i\oplus w^*\!$ ($1\leq i\leq N$). Then, the objective of the problem becomes to minimize $\max_{i\in [1,N]} \|\bar{w}^i\|$, and the constraint set of the problem becomes $\|w^i\oplus w^j\|=\|(w^i\oplus w^*)\oplus (w^j\oplus w^*)\|=\|\bar{w}^i\oplus \bar{w}^j\|\geq 2\delta_{BE}+1\;(\forall \; 1\!\leq\! i\!<\!j\!\leq\! N)$, so we reformulate the problem as follows.

\begin{equation}
\begin{aligned}
\arg\min_{\bar{w}^1,\dots,\bar{w}^N}&\max_{i\in [1,N]} \|\bar{w}^i\| \\
\text{s.t.}\quad &\|\bar{w}^i\!\oplus\! \bar{w}^j\|\!\geq\!2\delta_{BE}\!+\!1&&\! (\forall \; 1\!\leq\! i\!<\!j\!\leq\! N)\!\!\!\!\!
\end{aligned}
\end{equation}

The above problem does not involve $w^*$, so \textit{we solve it only once regardless of the dataset} to obtain $\bar{w}^i\;(1\leq i\leq N)$. To obtain $w^i\;(1\leq i\leq N)$ for a specific dataset, we first calculate $w^*$ with respect to this dataset, and then XOR $w^*$ and $\bar{w}^i$.

The above problem involves a min-max objective and ${N(N\!-\!1)}/{2}$ pairwise Hamming distance constraints, so it is difficult to find the optimal solution. Thus, we design a greedy algorithm, which, to simplify the objective $\max_{i\in[1,N]}\|\bar{w}^i\|$, generates $\bar{w}^i$ in increasing order of the Hamming weight $\|\bar{w}^i\|$, so that $\max_{i\in[1,N]}\|\bar{w}^i\|=\|\bar{w}^N\|$. Then, to minimize $\|\bar{w}^N\|$, $\|\bar{w}^i\|$ should increase as slowly as possible with the increase of $i$. Therefore, we gradually increase the Hamming weight from $0$, and for each Hamming weight, generate as many $\bar{w}^i$ as possible. To this end, we design a best-first search algorithm.

The algorithm initializes an empty set to store result bit strings and starts its traversal from $\mathbf{0}^L$, which has the smallest Hamming weight. To ensure that bit strings with smaller Hamming weights are added to the result set first, the algorithm keeps a priority queue of bit strings to be traversed, using Hamming weights as priorities. For each traversed bit string $\bar{w}$, the algorithm checks whether all Hamming distances between $\bar{w}$ and those in the result set are at least $2\delta_{BE}+1$. If so, $\bar{w}$ is added to the result set, and if the size of this set reaches $N$, the algorithm stops. Otherwise, the algorithm continues by iterating each bit string $\bar{w}'$ different by one bit from $\bar{w}$. If $\bar{w}'$ has not been traversed before, the algorithm marks $\bar{w}'$ as traversed and pushes $(\|\bar{w}'\|,\bar{w}')$ into the priority queue for subsequent traversals.

\subsection{Histogram-Based Watermark-Pattern Encoder}
\label{sec:overview:histogram}

To address the robustness issue discussed in Section~$\ref{subsec:watermark:workflow}$ for watermarked-histogram generation, we optimize the partial orders to create new constraints, which align with the partial orders but lead to larger gaps between the left-hand and right-hand cluster sizes in the watermarked histogram, thus making partial-order patterns resilient to fluctuations in cluster sizes. For example, using the partial order $x_{17}<x_{8}$ to constrain the watermarked histogram may result in $x_{17}-x_8=-1$, while the new constraints can result in $x_{17}-x_8=-10$. 

A naive method to generate these constraints is to introduce larger gaps into partial orders. For example, using $x_{17}<x_8-10$ instead of $x_{17}<x_8$ as a constraint. However, larger gaps also lead to worse utility, as they impose more restrictive constraints on the watermarked histogram, forcing it to deviate more from the original histogram. Therefore, it calls for a constraint-generation mechanism to enable the data owner to trade off robustness and utility. However, neither the quantitative relationship between gaps and robustness, nor that between gaps and utility is clear for the data owner to effectively determine the gaps. To address this issue, \remove{considering that it is infeasible to provide the data owner with control over all utility metrics that can support a wide range of applications, }we provide the data owner with quantitative control over robustness for this trade-off. Specifically, we parameterize robustness so that the data owner can express his/her robustness requirement by configuring a set of parameters. To meet the robustness requirement, we formulate a constrained optimization problem over the watermarked histogram, whose solution ensures the robustness requirement theoretically. The constraint generation mechanism will be detailed in Section~$\ref{sec:constraint_generator}$.

Since the constrained optimization problem is complex due to the complexity of meeting the robustness requirement, as we will see in Section~$\ref{sec:constraint:algorithm}$, directly solving the problem could lead to suboptimal watermarked histograms. For better utility, we design effective optimization mechanisms to create high-quality watermarked histograms, which will be described in Section~$\ref{sec:optimizer}$.

\begin{figure}[!t]
  \centering
  \includegraphics[width=\linewidth]{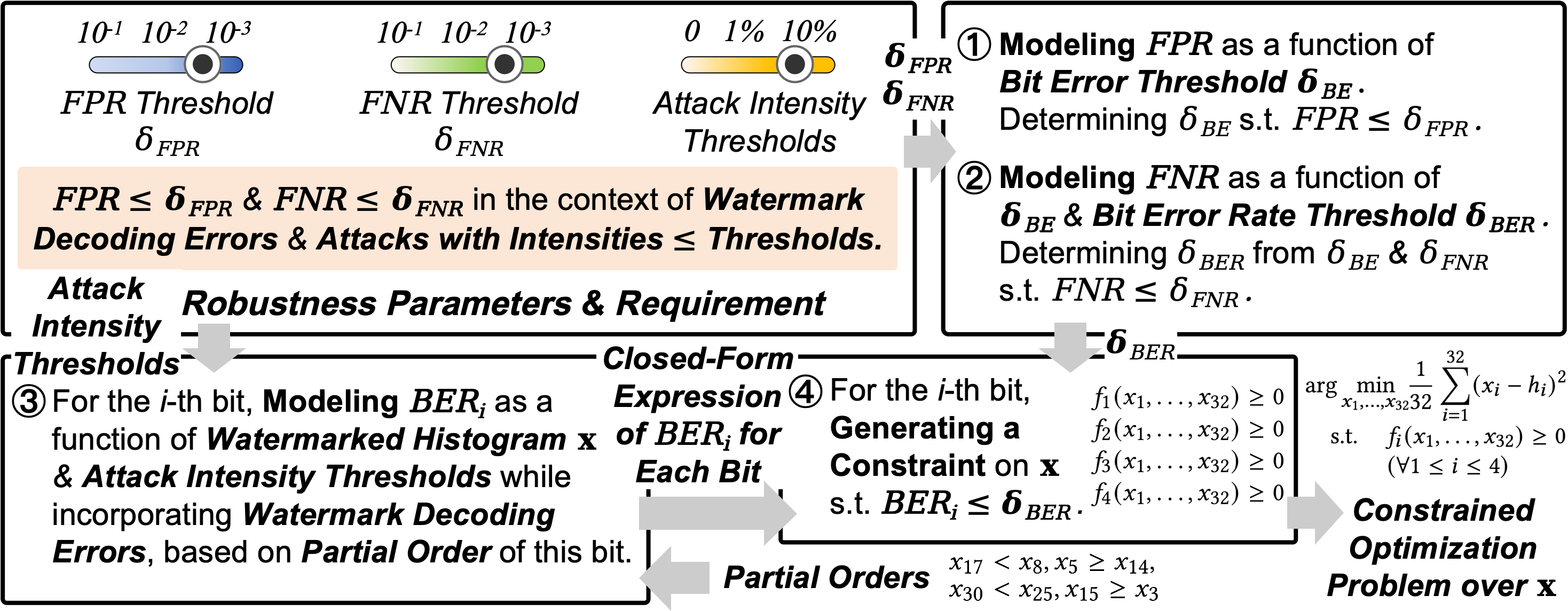}
  \caption{Overview of constraint generator.} \label{fig:constraints}
\end{figure}

\section{Constraint Generator for Watermarked Histograms}
\label{sec:constraint_generator}

To provide the data owner with effective control over robustness for robustness-utility trade-off, we present the constraint generator for watermarked-histogram generation. We first discuss the design principle in Section~\ref{sec:constraint:overview}, and then formulate a mathematical model to connect the data owner's robustness requirement with the watermarked histogram in Sections~\ref{sec:constraint:trace} and~\ref{sec:constraint:error_attack}. Based on this model, we formulate a constrained optimization problem over the watermarked histogram according to the robustness requirement in Section~\ref{sec:constraint:algorithm}. Figure~\ref{fig:constraints} shows an overview of the constraint generator.

\subsection{Constraint-Generator Overview} \label{sec:constraint:overview}

\hi{Design Principle.} To provide the data owner with effective control over robustness, we need to introduce a set of \myhi{robustness parameters} to allow the data owner to quantitatively express his/her \myhi{robustness requirement}. In particular, the robustness parameters refer to the threshold $\delta_{FPR}$ for the false positive rate (i.e., the probability of identifying a non-watermarked $D_s$ as watermarked), the threshold $\delta_{FNR}$ for the false negative rate (i.e., the probability of failing to identify the correct data buyer from a watermarked $D_s$), and the thresholds for intensities of a wide range of attacks, such as the proportion of deleted tuples in tuple-deletion attacks. Given a set of robustness parameters, the robustness requirement is expressed as follows: the false positive rate $FPR$ is below $\delta_{FPR}$ and the false negative rate $FNR$ is below $\delta_{FNR}$, in the context of potential \myhi{watermark decoding errors} (i.e., watermarks may be wrongly decoded even without attacks) and a set of \myhi{attacks} whose intensities do not exceed their corresponding thresholds. To meet the robustness requirement, we formulate a mathematical model to connect $FPR$, $FNR$, watermark decoding errors, and attack intensities with the watermarked histogram $\mathbf{x}$, so that by generating constraints on $\mathbf{x}$ based on this model and the robustness parameters, and formulating a constrained optimization problem over $\mathbf{x}$ accordingly, we theoretically ensure that the solution watermarked histogram satisfies the robustness requirement.

\hi{Formulating Mathematical Model.} We establish the mathematical model step by step. Firstly, we model $FPR$ as a function of $\delta_{BE}$ and determine $\delta_{BE}$ subject to $FPR\leq \delta_{FPR}$ to meet the requirement on $FPR$. Secondly, we model $FNR$ as a function of $\delta_{BE}$ and the threshold $\delta_{BER}$ for the bit error rate $BER_i$ (i.e., the probability that the $i$-th bit is wrongly decoded due to watermark decoding errors and attacks), where $BER_i\leq \delta_{BER}\;(1\!\leq\! i\!\leq\! L)$. We determine $\delta_{BER}$ subject to $FNR\!\leq\! \delta_{FNR}$ to meet the requirement on $FNR$. Thirdly, for the $i$-th bit, we model $BER_i$ as a function of $\mathbf{x}$ and attack intensity thresholds based on the partial order encoding this bit. Note that we also incorporate watermark decoding errors when modeling $BER_i$.

\hi{Formulating Constraint Optimization Problem.} For the $i$-th bit, we substitute $BER_i$ with its expression in terms of $\mathbf{x}$ and attack intensity thresholds in the inequality $BER_i\!\leq\! \delta_{BER}$ to satisfy the requirement on $FNR$ and generate a constraint on $\mathbf{x}$ (note that the requirement on $FPR$ is already satisfied by determining an appropriate $\delta_{BE}$ according to the mathematical model). For example, in Figure~\ref{fig:constraints}, we generate four constraints $f_i(x_1,\dots,x_{32})\!\geq\! 0\;(1\!\leq\! i\!\leq\! 4)$ that correspond to four bits encoded by four partial orders. Based on these constraints, we formulate a constrained optimization problem over $\mathbf{x}$, which will lead to a watermarked histogram that meets the robustness requirement.

\subsection{Modeling $FPR$ and $FNR$} \label{sec:constraint:trace}

\hi{Modeling $FPR$.} Identifying $D_s$ as watermarked means that if we input $D_s$ into the watermark decoding algorithm and obtain a bit string $\hat{w}$, then $\hat{w}$ matches a watermark in the database within $\delta_{BE}$ (i.e., $\exists i\in[1,\!N] \text{ s.t. }\|\hat{w}\oplus w^i\|\leq \delta_{BE}$). Moreover, under the condition that $D_s$ is not watermarked, $D_s$ could be any table other than the watermarked tables the data owner has sold, so we generally do not have information about $\hat{w}$ in this case. Therefore, according to the maximum entropy principle~\cite{MaximumEntropy}, we assume that $\hat{w}$ follows the $L$-bit uniform distribution $U[0,2^L\!-\!1]$, as in previous work~\cite{StableSignature,CycleGAN,DeepfakeFingerprint}, so we have

{
\begin{equation}
\label{eq:FPR1}
\begin{aligned}
\!FPR&=P\left(
\begin{aligned}
&\,\exists\, i\in[1,N]\; \text{ s.t. } \\[-4pt]
&\|\hat{w}\oplus w^i\|\leq \delta_{BE}
\end{aligned}
\,\middle|\,
\begin{aligned}
&\;\hat{w} \text{ is decoded from a} \\[-4pt]
&\text{ non-watermarked } D_s\\
\end{aligned}\right)\\
&=P\left(
\begin{aligned}
&\,\exists\, i\in[1,N]\; \text{ s.t. }\\[-4pt]
&\|\hat{w}\oplus w^i\|\leq \delta_{BE}
\end{aligned}\,\middle|\,\;\hat{w}\!\sim\! U[0,2^L\!-\!1]\right)\\[-21pt]
\end{aligned}
\end{equation}
}

Since our mechanism for watermark-database generation ensures that $\|w^i\oplus w^j\|\geq 2\delta_{BE}+1$ for $1\leq i<j\leq N$ (see Section~\ref{sec:overview:watermark_generator}), $\|\hat{w}\oplus w^i\|\leq \delta_{BE}\,(1\leq i\leq N)$ are disjoint events. In addition, due to the uniform distribution assumption of $\hat{w}$, these events have the same probability $\sum_{i=0}^{\delta_{BE}}\binom{L}{i}/2^L$, so we have

{
\setlength{\abovedisplayskip}{6pt}
\setlength{\belowdisplayskip}{6pt}
\begin{equation}
\label{eq:FPR2}
\begin{aligned}
FPR&=\sum_{i=1}^NP(\|\hat{w}\oplus w^i\|\leq \delta_{BE}\mid \hat{w}\sim U[0,2^L\!-\!1])\\&=\frac{N}{2^L}\sum_{i=0}^{\delta_{BE}}\binom{L}{i}
\end{aligned} \\[-5pt]
\end{equation}
}

Equation~\ref{eq:FPR2} shows that $FPR$ increases with $\delta_{BE}$, so we set $\delta_{BE}$ to the maximum subject to $FPR\leq \delta_{FPR}$, to maximize the bit error tolerance.

\hi{Modeling $FNR$.} Failing to identify the $i$-th data buyer from a watermarked $D_s$ that has been sold to this buyer means that $\hat{w}$ does not match $w^i$ (i.e., the watermark associated with the $i$-th data buyer and encoded in $D_s$) within $\delta_{BE}$ (i.e., $\|\hat{w}\oplus w^i\|\!>\! \delta_{BE}$), so we have

{
\setlength{\jot}{3pt}
\setlength{\abovedisplayskip}{6pt}
\setlength{\belowdisplayskip}{6pt}
\begin{equation}
\label{eq:FNR1}
\begin{aligned}
FNR&=P\left(\|\hat{w}\oplus w^i\|>\delta_{BE}\,\middle|\,
\begin{aligned}
&w^i\text{ is encoded in } D_s\,\text{,}\\[-4pt]
&\hat{w}\text{ is decoded from } D_s\\[-2pt]
\end{aligned}\right)
\end{aligned}
\end{equation}
}

As discussed in Section~\ref{sec:overview:cluster_pair_selector}, \oursys maximizes bit independence, ensuring that the decoding correctness of a single bit is weakly correlated with that of other bits, so we assume that bit errors are independent. Since the error rate of each bit may be different, it is difficult to formulate a closed-form expression of $FNR$, so we formulate an upper bound on $FNR$ that has a tractable expression. We observe that if the error rate of every bit is below $\delta_{BER}$, then the probability that $\|\hat{w}\oplus w^i\|\leq\delta_{BE}$ is above ${\sum_{j=0}^{\delta_{BE}}\binom{L}{j}\delta_{BER}^j(1-\delta_{BER})^{L-j}}$, and thus the probability that $\|\hat{w}\oplus w^i\|\!>\!\delta_{BE}$ is below ${1\!-\!\sum_{j=0}^{\delta_{BE}}\binom{L}{j}\delta_{BER}^j(1\!-\!\delta_{BER})^{L-j}}$. Therefore, we formulate this upper bound as below.

{
\setlength{\jot}{5pt}
\setlength{\abovedisplayskip}{6pt}
\setlength{\belowdisplayskip}{6pt}
\begin{equation}
\label{eq:FNR2}
\begin{aligned}
FNR&\leq1-\textstyle\sum_{j=0}^{\delta_{BE}}\binom{L}{j}\delta_{BER}^j(1-\delta_{BER})^{L-j},\\
&\;\;\;\;\;\text{if }BER_i\leq\delta_{BER}\;(\forall 1\leq i\leq L)
\end{aligned}
\end{equation}
}

Equation~\ref{eq:FNR2} shows that the upper bound on $FNR$ is a monotonically increasing function of $\delta_{BER}$. Therefore, we set $\delta_{BER}$ to the maximum subject to this upper bound below $\delta_{FNR}$. Then, by constraining $BER_i\leq \delta_{BER}\,(\forall 1\leq i\leq L)$, we ensure that $FNR\leq \delta_{FNR}$.

\subsection{Modeling Bit Error Rate} \label{sec:constraint:error_attack}

Since watermark decoding errors and attacks are potential causes of bit errors, we first model watermark decoding errors and attacks and then the bit error rate.

\hi{Modeling Watermark Decoding Errors and Attacks.} In \oursys, watermark decoding errors mean that during watermark decoding, tuples in $D_s$ may be assigned to wrong clusters (called cluster misassignments), which arise from limitations of the tuple embedding model. Specifically, during watermark encoding, the generated tuple latents are reconstructed into tuples to produce $D_w$, while during watermark decoding, tuples in $D_s$ are embedded into latents for cluster assignment (see Section~\ref{subsec:watermark:workflow}). However, the latents for the same tuple during watermark encoding and decoding are usually different due to the reconstruction error of the tuple embedding model, and thus may fall into different clusters.

We explicitly model three typical types of attacks: cell-level perturbation and alteration attacks that modify certain cells, and tuple-level deletion attacks that remove certain tuples. Each class of attacks is parameterized by an intensity, and we denote the intensity thresholds for perturbation, alteration, and tuple-deletion attacks as $I_{per}$, $I_{alt}$, and $I_{del}$, respectively. In perturbation attacks, the attacker injects noise into numerical attributes. Based on the fact that the attacker tends to inject larger noise into cells with larger absolute values and attribute standard deviations, we define the perturbation intensity as the ratio of noise standard deviations to the minimum of absolute cell values and attribute standard deviations. Moreover, we use Gaussian distributions to model noise following the maximum entropy principle~\cite{MaximumEntropy}, which implies that when there is no information about the noise distribution from the attacker, we choose the distribution with the maximum entropy to model noise, i.e., Gaussian distribution, given that the noise mean is nearly fixed (it is close to 0; otherwise the utility of the attacked table is seriously degraded), and the noise standard deviation is smaller than a fixed value (it is associated with the cell value and the attribute standard deviation that are fixed, and the perturbation intensity that is assumed below $I_{per}$ in the robustness requirement). We will show in experiments that \oursys is also resilient to other perturbations, such as uniform and Laplace noise. In alteration attacks, the attacker randomly changes a subset of categorical cells to random domain values, so we define the alteration intensity as the proportion of categorical cells the attacker modifies. In tuple-deletion attacks, the attacker deletes a subset of tuples, so we define the intensity of this class of attacks as the proportion of tuples the attacker deletes. Specifically, we consider adaptive tuple-deletion attacks, where the attacker performs clustering on the watermarked table, identifies $2L$ clusters that are likely selected to construct the watermark template as described in Section~\ref{sec:overview:cluster_pair_selector}, and uniformly deletes tuples from the $2L$ clusters (note that the cluster pairing relationships are hidden from the attacker). Although the attacker can delete tuples in a more structured way (e.g., selecting $L$ clusters from the $2L$ clusters and only deleting tuples from the $L$ clusters) to remove the watermark more effectively, these more structured attacks also introduce more systematic bias into the cluster distribution of the watermarked table, leading to a more severe utility loss. Therefore, we do not explicitly model these more structured attacks. We will show in experiments that in the context of these attacks, \oursys also achieves an effective robustness-utility trade-off. Note that tuple-insertion attacks are similar to tuple-deletion attacks in the sense that the insertion or deletion of a single tuple changes the size of a certain cluster by one, so by incorporating deletion attacks in the attack modeling \oursys also strikes a good balance between robustness and utility under insertion attacks, as we will see in experiments.

Watermark decoding errors, perturbation attacks, and alteration attacks result in cluster misassignments during watermark decoding, so we model watermark decoding errors and these attacks by modeling the cluster assignment process of individual tuples as probabilistic transitions between clusters captured by the matrix $\mathbf{P}=(p_{ij})_{1\leq i,j\leq M}$, where $p_{ij}$ is the probability that a tuple with the ground-truth cluster $C_i$ is assigned to cluster $C_j$, in the context of watermark decoding errors and these attacks. To estimate $\mathbf{P}$, during watermark-channel construction, after training the conditional latent diffusion model, we generate tuples in each cluster and simulate watermark decoding errors and attacks in the cluster assignment process of the generated tuples. Specifically, we first generate $T$ tuples in each cluster (called ground-truth cluster) as introduced in Section~\ref{subsec:watermark:workflow}. Next, we simulate perturbation and alteration attacks by adding Gaussian noise to numerical cells with noise standard deviations determined by $I_{per}$, as discussed above, and randomly altering categorical cells with probability $I_{alt}$. Then, we assign these tuples to clusters (called predicted clusters) as described in Section~\ref{subsec:watermark:workflow}, which involves watermark decoding errors. Finally, we compute a confusion matrix from the ground-truth and predicted clusters, and divide each entry in this matrix by its row sum to produce $\mathbf{P}$. Note that $T$ is a configurable parameter to balance estimation efficiency and the estimation error $\eta$, defined as the maximum absolute error between entries of the estimated and true transition matrices, and $\smash[t]{T\geq \frac{1}{2\eta^2}\ln{\frac{2M}{\xi}}}$ according to the Hoeffding inequality~\cite{Hoeffding}, where $\xi$ is the significance level that is typically set to $5\%$.

Usually, $\mathbf{P}$ is diagonally dominant, which means that tuples are assigned to their ground-truth clusters with high probabilities. Firstly, an effective tuple embedding model has a small reconstruction error, and the attacker does not excessively modify the watermarked table to avoid serious utility degradation, so the discrepancies between latents of the same tuple during watermark encoding and decoding are often small. Secondly, clustering algorithms tend to produce well-separated clusters, so these latent discrepancies perturb tuples across cluster boundaries with typically small probabilities.

To incorporate adaptive tuple-deletion attacks in the modeling, there are three key considerations. Firstly, deletions of different tuples are independent due to tuple independence of relational tables. Secondly, tuples within the $2L$ template clusters are often deleted with a higher probability than other tuples, due to the targeted nature of this class of attacks. Thirdly, tuples outside template clusters may still be deleted. Because $M$ is typically set to a large number, due to clustering randomness, the cluster structure produced by the data owner and that produced by the attacker are likely different. Therefore, we model adaptive tuple-deletion attacks by estimating deletion probabilities for tuples within/outside template clusters (denoted by $q_{in}$ and $q_{out}$, respectively). Specifically, we simulate adaptive tuple-deletion attacks 1000 times. In the $i$-th simulation, we perform clustering on the original table to form $M$ clusters using a unique random seed, identify $2L$ clusters from the clustering results as described in Section~\ref{sec:overview:cluster_pair_selector}, randomly delete $I_{del}\cdot\sum_{i=1}^Mh_i$ tuples from the $2L$ clusters, and count the number of tuples that fall into template clusters and are deleted, denoted by $n_i^{del}$. Eventually, $q_{in}\!=\!\frac{\max_{i=1}^{1000}n_i^{del}}{\sum_{i\in \{i|C_i\text{ is a template cluster}\}}h_i}$ and ${q_{out}\!=\!\frac{I_{del}\cdot\sum_{i=1}^Mh_i-\max_{i=1}^{1000}n_i^{del}}{\sum_{i\in \{i|C_i\text{ is not a template cluster}\}}h_i}}$, where for better robustness, we aggregate the simulation results using \texttt{max} to capture the worst case where deletion operations hit the most tuples in template clusters.

Putting it together, we model the process where $D_w$ is transformed into $D_s$ by attacks, and then tuples in $D_s$ are assigned to clusters for watermark decoding as follows: (1) Each tuple of cluster $C_k$ within $D_w$ remains in $D_s$ with probability $1-q_k$, where $q_k$ equals $q_{in}$ or $q_{out}$ depending on whether $C_k$ is a template cluster (w.r.t. adaptive tuple-deletion attacks). (2) If this tuple remains in $D_s$, it is probabilistically assigned to a cluster following the transition $\mathbf{P}$ (w.r.t. watermark decoding errors, perturbation attacks, and alteration attacks).

\hi{Modeling Bit Error Rate.} To model $BER_i$ for the $i$-th bit, by symmetry, we consider the bit error case where 0 is decoded as 1, and let the partial order encoding this bit be $x_{l_i}\geq x_{r_i}$. Moreover, we denote the extracted watermarked histogram as $\mathbf{y}=(y_j)_{j=1}^M$, which is a random vector due to the randomness of watermark decoding errors and attacks, with a distribution parameterized by the deterministic $\mathbf{x}$. Then, the bit error case means that $y_{l_i}<y_{r_i}$, so we have

{
\setlength{\jot}{1pt}
\setlength{\abovedisplayskip}{6pt}
\setlength{\belowdisplayskip}{6pt}
\begin{equation}
\label{eq:bit_abstract}
\begin{aligned}
BER_i=P(y_{l_i}<y_{r_i})=P(y_{l_i}-y_{r_i}<0)
\end{aligned}
\end{equation}
}

To model the distribution of $y_{l_i}-y_{r_i}$ for a closed-form expression of $BER_i$, we introduce the auxiliary random variables $u_{kl_i}$, $u_{kr_i}$, and $v_k$ for $1\leq k\leq M$, where $u_{kl_i}$ ($u_{kr_i}$) denotes the number of tuples generated in cluster $C_k$ and assigned to cluster $C_{l_i}\;(C_{r_i})$, $v_k$ denotes the number of tuples remaining in cluster $C_k$ after adaptive tuple-deletion attacks, and $u_{kl_i}\!\mid\! v_k$ ($u_{kr_i}\!\mid\! v_k$) denotes the conditional distribution of $u_{kl_i}$ ($u_{kr_i}$) given the value of $v_k$. Since $y_{l_i}$ ($y_{r_i}$) is the total number of tuples assigned to cluster $C_{l_i}\;(C_{r_i})$, we have

{
\setlength{\jot}{1pt}
\setlength{\abovedisplayskip}{6pt}
\setlength{\belowdisplayskip}{6pt}
\begin{equation}
\begin{aligned}
y_{j}&=\textstyle\sum_{k=1}^Mu_{kj}&&(j\in\{l_i,r_i\}) \\[-1pt]
\end{aligned}
\end{equation}
}

Based on our modeling of watermark decoding errors and attacks (i.e., a tuple of cluster $C_k$ within $D_w$ remains in $D_s$ with probability $1-q_k$, and if so, it is assigned to cluster $C_j$ with probability $p_{kj}$), we have

{
\setlength{\jot}{1pt}
\setlength{\abovedisplayskip}{6pt}
\setlength{\belowdisplayskip}{6pt}
\begin{equation}
\label{eq:uv}
\begin{aligned}
v_k&\sim B(x_k,1-q_k) \\
u_{kj}\mid v_k&\sim B(v_k, p_{kj}) && (j\in\{l_i,r_i\}) \\
u_{kj}&\sim B(x_k,(1-q_k)p_{kj}) && (j\in\{l_i,r_i\})
\end{aligned}
\end{equation}
}

We model $y_{l_i}$ and $y_{r_i}$ as normally distributed. To understand the reason, consider the distribution of $y_{l_i}$. Typically, $x_k\,(1\leq k\leq M)$ are large (i.e., from tens to hundreds), because the synthetic table usually has a large number of tuples to support complex applications (e.g., statistical analysis that involves high-dimensional aggregation queries). Moreover, $\mathbf{P}$ is diagonally dominant, as discussed before, so $x_k(1-q_k)p_{l_il_i}$ is often large. In addition, due to watermark decoding errors and attacks, $1-p_{l_il_i}$ (i.e., the cluster misassignment rate) is non-negligible and thus $1-(1-q_k)p_{l_il_i}$ is non-negligible, so $x_k(1-(1-q_k)p_{l_il_i})$ is relatively large. Therefore, we approximate the binomial random variable $u_{l_il_i}$ as normally distributed~\cite{TotalCovariance}. Furthermore, since diagonal entries of $\mathbf{P}$ account for a vast majority of the total probability mass, off-diagonal entries $p_{kl_i}\,(k\neq l_i)$ are usually negligible, which means that $(1-q_k)p_{kl_i}\,(k\neq l_i)$ are negligible and therefore $x_k(1-q_k)p_{kl_i}\,(k\neq l_i)$ are small, so we approximate the binomial random variables $u_{kl_i}\,(k\neq l_i)$ as Poisson distributed~\cite{TotalCovariance}. Since individual tuples undergo watermark decoding errors and attacks independently, $u_{kl_i}\,(1\leq k\leq M)$ are mutually independent. Thus, the sum $\sum_{k=1,k\neq l_i}^Mu_{kl_i}$ computed from multiple independent Poisson random variables also follows a Poisson distribution~\cite{TotalCovariance} with mean $\smash{\sum_{k=1,k\neq l_i}^M x_k(1-q_k)p_{kl_i}}$, which is usually large, so we further approximate this Poisson random variable as normally distributed~\cite{Possion}. Since $\smash[t]{\sum_{k=1,k\neq l_i}^Mu_{kl_i}}$ and $u_{l_il_i}$ are independent and have been approximated as normally distributed, we model $\smash[t]{y_{l_i}\!=\!\sum_{k=1,k\neq l_i}^Mu_{kl_i}\!+\!u_{l_il_i}}$ using a normal distribution. Similarly, we model $y_{r_i}$ as normally distributed.

Thus, $z_i\!:=\!y_{l_i}\!-\!y_{r_i}\!=\!\sum_{k=1}^M(u_{kl_i}\!-\!u_{kr_i})$ follows a normal distribution with the following mean and variance.

{
\setlength{\jot}{1pt}
\setlength{\abovedisplayskip}{6pt}
\setlength{\belowdisplayskip}{6pt}
\begin{equation}
\label{eq:ez_varz}
\begin{aligned}
\mathbb{E}[z_i]\!&=\!\textstyle\sum_{k=1}^M(\mathbb{E}[u_{kl_i}]-\mathbb{E}[u_{kr_i}]) \\
\!&=\!\textstyle\sum_{k=1}^Mx_k\!\cdot\! \underbrace{(1\!-\!q_k)(p_{kl_i}\!-\!p_{kr_i})}_{\quad\smash[t]{\alpha^{(i)}_k}} \\
\text{Var}(z_i)\!&=\!\textstyle\sum_{k=1}^M(\text{Var}(u_{kl_i})\!+\!\text{Var}(u_{kr_i})\!-\!2\text{Cov}(u_{kl_i},u_{kr_i})) \\
\!&=\!\textstyle\sum_{k=1}^Mx_k\cdot \underbrace{(1-q_k)(
\begin{aligned}[t]
&p_{kl_i}(1-(1-q_k)p_{kl_i})+ \\
&p_{kr_i}(1-(1-q_k)p_{kr_i})+ \\
&2(1-q_k)p_{kl_i} p_{kr_i})
\end{aligned}}_{\;\;\;\;\smash[t]{\beta^{(i)}_k}} \\[-8pt]
\end{aligned}
\end{equation}
}
where $\mathbb{E}[u_{kl_i}]$, $\mathbb{E}[u_{kr_i}]$, $\text{Var}(u_{kl_i})$, and $\text{Var}(u_{kr_i})$ come from Equation~\ref{eq:uv}, and $\text{Cov}(u_{kl_i}, u_{kr_i})$ is calculated by applying the law of total covariance~\cite{TotalCovariance} as follows.

{
\setlength{\jot}{1.5pt}
\setlength{\abovedisplayskip}{6pt}
\setlength{\belowdisplayskip}{6pt}
\begin{equation}
\begin{aligned}
\!\!\text{Cov}(u_{kl_i},\!u_{kr_i})&=\mathbb{E}[\text{Cov}(u_{kl_i}|v_k, u_{kr_i}|v_k)]+ \\
&\;\;\;\;\;\,\!\text{Cov}(\mathbb{E}[u_{kl_i}|v_k], \mathbb{E}[u_{kr_i}|v_k]) \\
&=\mathbb{E}[-v_k\!\cdot\! p_{kl_i} p_{kr_i}]+\text{Cov}(v_k\!\cdot\! p_{kl_i}, v_k\!\cdot\! p_{kr_i}) \\
&=-x_k\!\cdot\! (1-q_k)^2p_{kl_i} p_{kr_i} \qquad (1\leq k\leq M)
\end{aligned}
\end{equation}
}

Therefore, $BER_i\!=\!P(z_i\!<\!0)\!=\!\Phi({-\mathbb{E}[z_i]}/{\sqrt{\text{Var}(z_i)}})$, where $\Phi$ is the cumulative distribution function of the standard normal distribution, when the bit error case means that 0 is decoded as 1 (the case discussed so far) and the partial order encoding this bit is $x_{l_i}\!\geq\! x_{r_i}$. Similarly, when the bit error case means that 1 is decoded as 0 and the underlying partial order is $x_{l_i}\!<\!x_{r_i}$, $BER_i\!=\!\Phi({\mathbb{E}[z_i]}/{\sqrt{\text{Var}(z_i)}})$, so we have

{
\setlength{\abovedisplayskip}{6pt}
\setlength{\belowdisplayskip}{6pt}
\begin{equation}
\label{eq:bit}
\begin{aligned}
\!\!BER_i\!=\!
\begin{cases}
\Phi(\frac{-\mathbb{E}[z_i]}{\sqrt{\text{Var}(z_i)}}), & \!\!\!\text{if partial order is }x_{l_i}\!\geq\! x_{r_i} \\
\Phi(\frac{\mathbb{E}[z_i]}{\sqrt{\text{Var}(z_i)}}), & \!\!\!\text{if partial order is }x_{l_i}\!<\! x_{r_i}
\end{cases}
\end{aligned}
\end{equation}
}

\subsection{Constrained Optimization Problem over $\mathbf{x}$} \label{sec:constraint:algorithm}

For the $i$-th bit, by substituting $BER_i$ with its closed-form expression (see Equation~\ref{eq:bit}) in the inequality $BER_i\leq \delta_{BER}$, and replacing $\mathbb{E}[z_i]$ and $\text{Var}(z_i)$ with $\sum_{k=1}^M\!x_k\alpha^{(i)}_k$ and $\sum_{k=1}^M\!x_k\beta^{(i)}_k$, respectively ($\alpha^{(i)}_k$ and $\smash{\beta^{(i)}_k}$ are determined by attack intensity thresholds; see Equation~\ref{eq:ez_varz}), we have the following constraints for this bit.

{
\setlength{\jot}{1pt}
\setlength{\abovedisplayskip}{6pt}
\setlength{\belowdisplayskip}{6pt}
\begin{equation}
\label{eq:xlr}
\begin{aligned}
&\hspace{-22.75em}x_{l_i}\mathrel{\substack{\geq \\ <}} x_{r_i}\xRightarrow{\text{}} \\
\;\;\;\begin{cases}
\begin{aligned}
\textstyle\sum_{k=1}^M\!x_k \alpha_k^{(i)}\mathrel{\substack{\geq \\ <}} 0 \\[3pt]
(\textstyle\sum_{k=1}^M\!x_k \alpha_k^{(i)})^2\!-\!\Phi^{-1}(\delta_{BER})^2 \textstyle\sum_{k=1}^M\!x_k \beta_k^{(i)}\geq 0
\end{aligned}
\end{cases}
\end{aligned}
\end{equation}
}
where $l_i$ and $r_i$ are cluster ordinals of the partial order that encodes the $i$-th bit. Based on these constraints, we formulate a constrained optimization problem over $\mathbf{x}$ as follows.

\begin{equation}
\label{eq:problem}
\begin{aligned}
\!\!\!\!\!\!\arg&\min_{x_1,\dots,x_M} \frac{1}{M}\textstyle\sum_{i=1}^M(x_i-h_i)^2 \\
\text{s.t.}\quad\!\!\!\!&\textstyle\sum_{k=1}^Mx_k \alpha_k^{(i)} 
\begin{cases}
\begin{aligned}
\geq 0, & \text{ if } w_i=0 \\
< 0, & \text{ if } w_i=1 \\[-2pt]
\end{aligned}
\end{cases} \qquad\quad (\forall 1\leq i\leq L)\\[3pt]
&\!(\textstyle\sum_{k=1}^Mx_k \alpha_k^{(i)})^2-\Phi^{-1}(\delta_{BER})^2 \textstyle\sum_{k=1}^Mx_k \beta_k^{(i)}\geq 0  \\
&\qquad\qquad\qquad\qquad\qquad\qquad\qquad\qquad\;\;\, (\forall 1\leq i\leq L)\\[3pt]
&\textstyle\sum_{i=1}^Mx_i=\textstyle\sum_{i=1}^Mh_i\;\;(x_1,\dots,x_M \in \mathbb{N^+})
\end{aligned}
\end{equation}
where $w$ is the $L$-bit watermark to be encoded, the objective is the mean squared error between the watermarked and original histograms (we use this objective because it imposes heavier penalties on larger deviations, leading to better utility), the first and second sets of constraints correspond to the generated constraints, and the third constraint requires that $D_w$ and $D_o$ have the same table cardinality, which is crucial for certain downstream tasks of $D_w$ (e.g., counting queries).

This problem is an integer programming problem since $\mathbf{x}$ are integers. In addition, this problem is non-convex because the feasible set defined by the second set of constraints is non-convex. Moreover, the terms $\smash{\sum_{k=1}^Mx_k \alpha_k^{(i)}}$ and $\smash{\sum_{k=1}^Mx_k \beta_k^{(i)}}$ in constraints involve a large number of variables, leading to a highly entangled problem structure. Therefore, directly solving this problem usually results in a suboptimal solution, so we subsequently present our optimization mechanisms that lead to a better solution in Section~\ref{sec:optimizer}.

\section{Optimizer for Watermarked Histograms} \label{sec:optimizer}

To solve the constrained optimization problem in Section~\ref{sec:constraint:algorithm}, we present the constraint simplification mechanism in Section~\ref{sec:optimizer:simplifier} and the multi-stage optimization mechanism in Section~\ref{sec:optimizer:multi}.

\subsection{Constraint Simplification Mechanism} \label{sec:optimizer:simplifier}

The general idea to simplify constraints is to substitute complex terms with simpler upper or lower bounds, so that the simplified constraints align with the original ones but the number of variables in each constraint is reduced, leading to a tractable optimization problem. However, this leads to a challenge: We need to determine suitable bounds that do not excessively tighten the original constraints. Otherwise, the solution space may be significantly reduced, potentially excluding high-quality solutions. To address this challenge, we first determine an initial feasible set of $\mathbf{x}$, which is expressed by simple constraints and is close to the original solution space. We then calculate surrogate bounds on complex terms from this set by solving simple constrained optimization problems. Finally, the constraints of the simplified problem include (1) constraints that express the initial feasible set of $\mathbf{x}$, (2) constraints whose complex terms are substituted with simpler bounds, and (3) other constraints from the original problem.

To determine the initial feasible set of $\mathbf{x}$, we observe that a high-quality watermarked histogram should deviate marginally from the original histogram, so $\mathbf{x}$ can be safely bounded in a small interval centered at $\mathbf{h}$. To control the bound tightness, we introduce $\tau\in [0,1]$ and bound $x_i$ in $\smash{[h_i-\tau\cdot\sum^M_{j=1}h_j,h_i+\tau\cdot\sum^M_{j=1}h_j]}$ for $1\!\leq\! i\!\leq\! M$, where $\smash{\sum^M_{j=1}h_j}$ is the table cardinality. The configuration of $\tau$ will be discussed in Section~\ref{sec:optimizer:multi}. In addition, since the constraints of the original problem align with partial orders, we further restrict the initial feasible set of $\mathbf{x}$ by these simple partial orders. Moreover, we restrict this set by the simple equal-cardinality constraint from the original problem.

To derive surrogate bounds on the complex terms to be substituted, consider the term $\smash{\sum_{k=1}^Mx_k \alpha_k^{(i)}}$, there are two key considerations. Firstly, the dominant terms of $\smash{\sum_{k=1}^Mx_k \alpha_k^{(i)}}$ are $\smash{x_{l_i}\alpha_{l_i}^{(i)}}$ and $\smash{x_{r_i}\alpha_{r_i}^{(i)}}$, because the transition matrix $\mathbf{P}$ is diagonally dominant (see Section~$\ref{sec:constraint:error_attack}$), while $\smash{x_{l_i}\alpha_{l_i}^{(i)}}$ and $\smash{x_{r_i}\alpha_{r_i}^{(i)}}$ are the only terms that involve diagonal entries of $\mathbf{P}$ (see Equation~\ref{eq:ez_varz}). Thus, the target bound should involve only $x_{l_i}$ and $x_{r_i}$. Secondly, since $\smash{(\sum_{k=1}^Mx_k \alpha_k^{(i)})^2}$ needs to be lower bounded (i.e., substitute this term with its lower bound) based on the second set of constraints in Equation~$\ref{eq:problem}$, when $w_i=0$ and $\smash[t]{\sum_{k=1}^Mx_k \alpha_k^{(i)}\geq0}$ (see Equation~\ref{eq:xlr}), $\smash[t]{\sum_{k=1}^Mx_k \alpha_k^{(i)}}$ needs to be lower bounded. Similarly, when $w_i=1$ and $\smash[b]{\sum_{k=1}^Mx_k \alpha_k^{(i)}<0}$, $\smash{\sum_{k=1}^Mx_k \alpha_k^{(i)}}$ needs to be upper bounded. Therefore, the target bound can be expressed as $\smash{x_{l_i} \alpha_{l_i}^{(i)}+x_{r_i} \alpha_{r_i}^{(i)}+E_i}$, where $E_i$ is the result of the following linear problem that can be solved efficiently.

\vspace{-1em}
\begin{equation}
\label{eq:simple}
\begin{aligned}
&\!\!\!\!\!\!\!\!\!\!\!\begin{cases}
\displaystyle{\min_{x_1,\dots,x_M}} &\sum_{k=1,k \notin \{l_i, r_i\}}^Mx_k \alpha^{(i)}_k, \text{ if } w_i=0 \\[-2pt]
\displaystyle{\max_{x_1,\dots,x_M}} &\sum_{k=1,k \notin \{l_i, r_i\}}^Mx_k \alpha^{(i)}_k, \text{ if } w_i=1 \\[1pt]
\end{cases} \\
\text{s.t.} \quad\!\!\!&|x_j-h_j|\leq\tau\cdot \textstyle\sum^M_{k=1}h_k&&\hspace{-4.85em} (\forall 1\leq j\leq M)\hspace{-1em} \\
&x_{l_j}-x_{r_j}
\begin{cases}
\begin{aligned}
\geq 0, & \text{ if }w_j=0\\
<0, & \text{ if }w_j=1\\[-2pt]
\end{aligned}
\end{cases}
&&\hspace{-4.8em} (\forall 1\leq j\leq L)\hspace{-1em} \\
&\textstyle\sum^M_{j=1}x_j=\textstyle\sum^M_{j=1}h_j\;\;(x_1,\dots,x_M \in \mathbb{N^+})\hspace{-1em}
\end{aligned}
\end{equation}
where the first set of constraints is the bounds on $\mathbf{x}$, the second set of constraints is partial orders, and the third is the equal-cardinality constraint. Similarly, we can calculate a surrogate bound on $\smash[t]{\sum_{k=1}^Mx_k\beta_k^{(i)}}$.

\subsection{Multi-Stage Optimization Mechanism}
\label{sec:optimizer:multi}

In the constraint simplification mechanism, $\tau$ determines the bounds on $\mathbf{x}$, and an inappropriate choice of $\tau$ may significantly shrink the solution space of the simplified problem relative to the original problem, potentially resulting in a suboptimal solution. If the bounds on $\mathbf{x}$ are too loose, the surrogate bounds used to replace complex terms also become loose, since they are obtained by solving constrained optimization problems in which the bounds on $\mathbf{x}$ are part of the constraints. Consequently, the original constraints may become overly restrictive after substitution. Conversely, if the bounds on $\mathbf{x}$ are overly tight, the solution space is directly reduced, as these bounds are explicitly embedded in the constraints of the simplified problem.

To address this issue, we adaptively adjust $\tau$ based on intermediate values of $\mathbf{x}$, ensuring that the bounds on $\mathbf{x}$ always align with the current solution space. Specifically, we divide the optimization process into multiple stages, initialize $\tau$ at the beginning, and adjust $\tau$ to $\smash{\max_{1\leq i\leq M}|x_i-h_i|/\sum^M_{j=1}h_j}$ at the end of each stage.

\section{Experiments} \label{sec:experiments}

In this section, we present an extensive evaluation of \oursys to answer the following research questions.

\begin{enumerate}[label=\textbf{RQ\arabic*)}, leftmargin=32pt]
\item Does \oursys achieve effective traceability for tracing a large number of data buyers, strong robustness against attacks, and high utility in terms of distribution similarity to the original table, and performance of machine learning and statistical analysis? (see Section~\ref{sec:experiments:robustness_utility})
\item Does the constraint generation mechanism provide an effective trade-off between robustness and utility? (see Section~\ref{sec:experiments:constraint})
\item Does the optimization mechanisms effectively improve utility? (see Section~\ref{sec:experiments:optimizer})
\item What are the trade-offs among the hyperparameters $M$, $N$, and $L$? (see Section~\ref{sec:experiments:mnl})
\item What is the runtime performance of \oursys? (see Section~\ref{sec:experiments:time})
\end{enumerate}

\subsection{Experimental Settings}

\subsubsection{Environments}

All experiments were conducted on an Ubuntu 24.04.1 LTS server, equipped with a 3.10 GHz Intel Xeon Gold 6242R CPU with 40 physical cores, 256 GB RAM, and an NVIDIA RTX 3090 GPU with 24 GB VRAM.

\subsubsection{Datasets}

We evaluated \oursys on four real-world datasets from the UCI Machine Learning Repository~\cite{UCI}: Beijing, Electric, Phishing, and Shoppers, each associated with a classification or regression task. We split these datasets into training, validation, and test sets in an 8:1:1 ratio following~\cite{TabSyn}, using a fixed random seed for reproducibility. Then, the original table used to train the table synthesis models referred to the training and validation sets. Table~$\ref{tab:experiments:dataset}$ lists statistics of these datasets.

\subsubsection{Metrics}

We used the following metrics, where traceability accuracy measured traceability and robustness, and the other metrics measured utility. For utility evaluation, the marginal distribution gap and correlation gap measured the distribution difference between the synthetic table (either watermarked or non-watermarked) and the original table, the machine learning efficacy gap measured the machine learning performance of the synthetic table, and the relative aggregation query error measured the statistical analysis performance of the synthetic table.

\setlength{\tabcolsep}{3.5pt}
\begin{table}[t]
\centering
\caption{Statistics of datasets used for evaluation.}
\label{tab:experiments:dataset}
\renewcommand{\arraystretch}{0.9}
{\fontfamily{ntxtlf}\fontsize{9pt}{11.5pt}\selectfont
\begin{tabular}{l c c c c c c}
\toprule
Name & \# Train & \# Val & \# Test & \# Num & \# Cat & ML Tasks \\
\midrule
Beijing & 33405 & 4176 & 4176 & 7 & 5 & Regression \\
Electric & 8000 & 1000 & 1000 & 12 & 1 & Classification \\
Phishing & 8843 & 1106 & 1106 & 0 & 31 & Classification \\
Shoppers & 9864 & 1233 & 1233 & 10 & 8 & Classification \\
\bottomrule
\end{tabular}
}
\end{table}

\noindent{$\bullet$\textit{Traceability Accuracy (Acc.)}} was calculated by dividing the number of trials where the data buyer was correctly identified by the total number of trials.

\noindent{$\bullet$\textit{Marginal Distribution Gap (Marg. Gap)}} was the univariate distribution difference between the synthetic and original tables, measured using the Kolmogorov-Smirnov test and the total variation distance for numerical and categorical attributes, respectively.

\noindent{$\bullet$\textit{Correlation Gap (Corr. Gap)}} was the pairwise distribution difference between the synthetic and original tables, measured using the Pearson correlation coefficient and the total variation distance for numerical and categorical attributes, respectively.

\noindent{$\bullet$\textit{Machine Learning Efficacy Gap (MLE Gap)}} measured the degradation in machine learning performance (in terms of AUROC and RMSE, for classification and regression tasks, respectively) of the synthetic table relative to the original table. To evaluate MLE Gap, we trained two XGBoost models using the synthetic and original tables, respectively, evaluated their performance using the test set of the dataset, and calculated the performance gap between these models.

\noindent{$\bullet$\textit{Relative Aggregation Query Error (RAQE)}} was the relative error between aggregation query results of the synthetic and original tables. Following existing work on density estimation~\cite{CE1,CE2,CE3,CE4}, for each dataset, we generated 3000 queries for each aggregation function, including \texttt{COUNT} and \texttt{AVG}. We generated each query in four steps: (1) Randomly selected a numerical column to place the aggregation function. (2) Randomly selected the number of predicates $f$ from [1, \#attributes]. (3) Randomly selected $f$ distinct columns to place the predicates. For numerical columns, the predicates were uniformly selected from $\{\leq,\geq\}$. For categorical columns, only equality predicates were selected. (4) Randomly selected a tuple from the original table and used its attributes as predicate literals. In addition, we randomly selected 10\% of the queries with categorical predicates and changed one of the categorical predicates to the \texttt{GROUP} \texttt{BY} predicate. To simulate different query patterns, we retained queries whose predicate selectivities were concentrated around three representative selectivities: 1\% (low), 5\% (medium), and 20\% (high), and ensured that each selectivity corresponded to one third of the queries (i.e., 1000 queries per selectivity per aggregation function). Note that Phishing did not have numerical attributes, so we did not generate \texttt{AVG} queries for this dataset (denoted by N/A in Table~\ref{tab:experiments:quality_detection}).

\subsubsection{Baselines}

We compared \oursys with the following baselines.

\noindent{$\bullet$\textit{w/o WM}} was a non-watermarked baseline used for utility comparison.

\noindent{$\bullet$\textit{TabularMark}}~\cite{TabularMark} was a recent scheme that encoded the watermark by first synthesizing a non-watermarked table, and then injecting large-scale noise of specific patterns into a moderate proportion of cells (called watermarked cells) within this table. This design made TabularMark limited in utility, because large-scale noise could compromise tuple integrity, especially when watermarks were encoded in categorical attributes, as we will see in our experiments. In contrast, \oursys ensured tuple integrity by first determining a histogram, and then synthesizing a watermarked table according to this histogram, without manipulating individual tuples. For better utility, we implemented TabularMark by encoding watermarks into numerical attributes for all datasets except for Phishing, a categorical dataset. To extend TabularMark to a multi-bit scheme, we encoded each bit into distinct watermarked cells.

\noindent{$\bullet$\textit{FreqyWM}}~\cite{FreqWM} was a recent scheme that encoded the watermark by synthesizing a non-watermarked table and forming specific congruence-relation patterns (i.e., equivalences between integers based on whether they were equal modulo some integer) between partition sizes of this table, via duplicating or removing certain tuples. Since congruence relations were sensitive to small fluctuations in partition sizes, FreqyWM was limited in robustness. In contrast, \oursys designed partial-order patterns that were more resilient. To extend FreqyWM to a multi-bit scheme, we encoded each bit into a distinct congruence relation.

\noindent{$\bullet$\textit{TabWak}}~\cite{TabWak} was a recent scheme tailored for synthetic tabular data. It encoded the watermark into Gaussian noise vectors, synthesized the watermarked table from these vectors using DDIM~\cite{DDIM} (an invertible diffusion model), and exploited the invertibility of this diffusion model to recover noise vectors from tuples for watermark decoding. To support multi-bit watermarks, we vertically partitioned noise vectors into multiple segments and encoded each bit into the same watermark pattern as TabWak within a distinct segment.

\subsubsection{Attacks for Robustness Evaluation}

We considered the following attacks:

\noindent{$\bullet$\textit{Perturbation}}: The attacker added noise to numerical cells. We considered three typical noise distributions: Gaussian (Gau.), uniform (Uni.), and Laplace (Lap.). Since Phishing did not have numerical attributes, we did not launch perturbation attacks on this dataset (denoted by N/A in Table~\ref{tab:experiments:quality_detection}). The standard deviation of the noise added to each numerical cell was equal to the attack intensity multiplied by the minimum of the absolute cell value and the attribute standard deviation.

\noindent{$\bullet$\textit{Alteration (Alt.)}}: The attacker randomly modified a proportion of categorical cells within the watermarked table. The intensity of this class of attacks was defined as the proportion of modified categorical cells.

\noindent{$\bullet$\textit{Tuple deletion (Del.)}}: The attacker randomly deleted a subset of tuples from the watermarked table. The intensity of tuple-deletion attacks was defined as the proportion of tuples the attacker deleted.

\noindent{$\bullet$\textit{Tuple insertion (Ins.)}}: The attacker randomly duplicated a subset of existing tuples and inserted them into the watermarked table. Similarly, the intensity of these attacks was defined as the proportion of inserted tuples.

\noindent{$\bullet$\textit{Regeneration (Gen.)}}: The attacker trained a tuple embedding model on the watermarked table, mapped tuples to latents, added Gaussian noise to tuple latents, and regenerated a table from the noisy tuple latents using the tuple embedding model. The standard deviation of the noise added to each latent element was equal to the attack intensity multiplied by the latent element.

\noindent{$\bullet$\textit{Adaptive tuple deletion (Ada.)}}: The attacker first performed clustering on the watermarked table to form $M$ clusters, and then identified $2L$ clusters that were likely selected to construct the watermark template according to the cluster histogram, as described in Section~\ref{sec:overview:cluster_pair_selector}. Since the cluster pairing relationships were hidden from the attacker by the pseudo-random pairing strategy, we simulated this class of attacks by randomly selecting $L$ clusters from the $2L$ clusters and uniformly deleting tuples from the $L$ clusters. The intensity of adaptive tuple-deletion attacks was defined as the proportion of tuples the attacker deleted from the watermarked table.

\noindent{$\bullet$\textit{Shadow model (Sha.)}}: The attacker trained a tuple embedding model and a conditional latent diffusion model (called shadow models) on the watermarked table to synthesize a new table. We did not expect \oursys and all baselines to resist shadow-model attacks (the traceability accuracy of all schemes was 0 under shadow-model attacks; see Table~\ref{tab:experiments:quality_detection}), because these attacks would seriously degrade utility as table synthesis inevitably caused utility loss and resynthesizing a synthetic table magnified utility loss. To validate this claim, we measured the utility in terms of RAQE and MLE Gap of the watermarked table under various attacks (note that in the case of deletion attacks, the \texttt{COUNT} query results were scaled by the cardinality of the attacked table when computing RAQE to prevent overstating the utility loss). The results are shown in Tables~\ref{tab:experiments:degrade_beijing}-\!\!~\ref{tab:experiments:degrade_shoppers}, where shadow-model attacks constantly led to a relative utility degradation exceeding 30\% compared to the watermarked table without attacks across all datasets, even 50\% on Beijing and Shoppers.

We set the intensity to 5\% for all attacks, except for shadow-model attacks that were not parameterized by an intensity.

\setlength{\tabcolsep}{2.7pt}
\begin{table}[!t]
\caption{Utility in terms of RAQE (with query cardinality scaling for \texttt{COUNT} in the presence of deletion attacks) and MLE Gap under different attacks on Beijing. The underlying watermarking scheme is \oursys.}
\label{tab:experiments:degrade_beijing}
\renewcommand{\arraystretch}{0.85}
{\fontfamily{ntxtlf}\fontsize{9pt}{11.5pt}\selectfont
\begin{tabular}{l||c c c c c c c c}
\toprule
\multicolumn{1}{c||}{Attacks} & \makecell{\texttt{COUNT}\\$@1\%$} & \makecell{\texttt{AVG}\\$@1\%$} & \makecell{\texttt{COUNT}\\$@5\%$} & \makecell{\texttt{AVG}\\$@5\%$} & \makecell{\texttt{COUNT}\\\!$@20\%$\!} & \makecell{\texttt{AVG}\\\!$@20\%$\!} & \makecell{MLE\\Gap} \\
\midrule
\,\,w/o\, Att. & $0.232$ & $0.727$ & $0.172$ & $0.342$ & $0.059$ & $0.308$ & $0.108$ \\
\midrule
\;\,5\%\, Gau. & $0.404$ & $0.767$ & $0.272$ & $0.366$ & $0.097$ & $0.316$ & $0.113$ \\
10\%\, Gau. & $0.375$ & $0.769$ & $0.259$ & $0.363$ & $0.094$ & $0.316$ & $0.119$ \\
\midrule
\;\,5\%\, Alt. & $0.222$ & $0.759$ & $0.168$ & $0.401$ & $0.059$ & $0.304$ & $0.124$ \\
10\%\, Alt. & $0.228$ & $0.836$ & $0.172$ & $0.486$ & $0.061$ & $0.304$ & $0.136$ \\
\midrule
\;\,5\%\, Del. & $0.234$ & $0.743$ & $0.171$ & $0.344$ & $0.060$ & $0.304$ & $0.111$ \\
10\%\, Del. & $0.235$ & $0.745$ & $0.172$ & $0.350$ & $0.060$ & $0.309$ & $0.115$ \\
\midrule
\;\,5\%\, Gen. & $0.208$ & $0.846$ & $0.147$ & $0.365$ & $0.037$ & $0.305$ & $0.118$ \\
10\%\, Gen. & $0.217$ & $1.021$ & $0.141$ & $0.497$ & $0.039$ & $0.426$ & $0.141$ \\
\midrule
\;\,5\%\, Ada. & $0.242$ & $0.751$ & $0.180$ & $0.369$ & $0.068$ & $0.372$ & $0.112$ \\
10\%\, Ada. & $0.274$ & $0.799$ & $0.203$ & $0.440$ & $0.083$ & $0.484$ & $0.120$ \\
\midrule
\,\;\,\;\;\;\;\;\;Sha. & $0.478$ & $1.235$ & $0.385$ & $0.846$ & $0.180$ & $0.904$ & $0.223$ \\
\bottomrule
\end{tabular}
}
\end{table}

\setlength{\tabcolsep}{2.5pt}
\begin{table}[!t]
\caption{Utility in terms of RAQE (with query cardinality scaling for \texttt{COUNT} in the presence of deletion attacks) and MLE Gap under different attacks on Electric. The underlying watermarking scheme is \oursys.}
\label{tab:experiments:degrade_electric}
\renewcommand{\arraystretch}{0.85}
{\fontfamily{ntxtlf}\fontsize{9pt}{11.5pt}\selectfont
\begin{tabular}{l||c c c c c c c c}
\toprule
\multicolumn{1}{c||}{Attacks} & \makecell{\texttt{COUNT}\\$@1\%$} & \makecell{\texttt{AVG}\\$@1\%$} & \makecell{\texttt{COUNT}\\$@5\%$} & \makecell{\texttt{AVG}\\$@5\%$} & \makecell{\texttt{COUNT}\\\!$@20\%$\!} & \makecell{\texttt{AVG}\\\!$@20\%$\!} & \makecell{MLE\\Gap} \\
\midrule
\,\,w/o\, Att. & $0.261$ & $0.118$ & $0.118$ & $0.050$ & $0.055$ & $0.023$ & $0.017$ \\
\midrule
\;\,5\%\, Gau. & $0.259$ & $0.119$ & $0.116$ & $0.050$ & $0.054$ & $0.023$ & $0.017$ \\
10\%\, Gau. & $0.264$ & $0.118$ & $0.112$ & $0.050$ & $0.052$ & $0.023$ & $0.017$ \\
\midrule
\;\,5\%\, Alt. & $0.267$ & $0.119$ & $0.121$ & $0.051$ & $0.056$ & $0.024$ & $0.022$ \\
10\%\, Alt. & $0.276$ & $0.121$ & $0.129$ & $0.053$ & $0.063$ & $0.027$ & $0.027$ \\
\midrule
\;\,5\%\, Del. & $0.264$ & $0.121$ & $0.119$ & $0.051$ & $0.054$ & $0.023$ & $0.017$ \\
10\%\, Del. & $0.268$ & $0.123$ & $0.121$ & $0.052$ & $0.055$ & $0.023$ & $0.016$ \\
\midrule
\;\,5\%\, Gen. & $0.261$ & $0.119$ & $0.120$ & $0.051$ & $0.056$ & $0.023$ & $0.017$ \\
10\%\, Gen. & $0.268$ & $0.119$ & $0.124$ & $0.051$ & $0.060$ & $0.023$ & $0.018$ \\
\midrule
\;\,5\%\, Ada. & $0.271$ & $0.122$ & $0.127$ & $0.053$ & $0.059$ & $0.025$ & $0.017$ \\
10\%\, Ada. & $0.297$ & $0.129$ & $0.146$ & $0.059$ & $0.071$ & $0.031$ & $0.018$ \\
\midrule
\,\;\,\;\;\;\;\;\;Sha. & $0.308$ & $0.135$ & $0.148$ & $0.061$ & $0.069$ & $0.029$ & $0.024$ \\
\bottomrule
\end{tabular}
}
\end{table}

\setlength{\tabcolsep}{2.7pt}
\begin{table}[!t]
\caption{Utility in terms of RAQE (with query cardinality scaling for \texttt{COUNT} in the presence of deletion attacks) and MLE Gap under different attacks on Phishing. The underlying watermarking scheme is \oursys.}
\label{tab:experiments:degrade_phishing}
\centering
\renewcommand{\arraystretch}{0.85}
{\fontfamily{ntxtlf}\fontsize{9pt}{11.5pt}\selectfont
\begin{tabular}{l||c c c c c}
\toprule
\multicolumn{1}{c||}{Attacks} & \makecell{\texttt{COUNT}\\$@1\%$} & \makecell{\texttt{COUNT}\\$@5\%$} & \makecell{\texttt{COUNT}\\$@20\%$} & \makecell{MLE\\Gap} \\
\midrule
\,\,w/o\, Att. & $0.276$ & $0.120$ & $0.060$ & $0.008$ \\
\midrule
\;\,5\%\, Alt. & $0.458$ & $0.300$ & $0.189$ & $0.011$ \\
10\%\, Alt. & $0.669$ & $0.524$ & $0.337$ & $0.013$ \\
\midrule
\;\,5\%\, Del. & $0.280$ & $0.122$ & $0.060$ & $0.009$ \\
10\%\, Del. & $0.283$ & $0.124$ & $0.061$ & $0.008$ \\
\midrule
\;\,5\%\, Gen. & $0.276$ & $0.121$ & $0.060$ & $0.008$ \\
10\%\, Gen. & $0.276$ & $0.121$ & $0.060$ & $0.008$ \\
\midrule
\;\,5\%\, Ada. & $0.301$ & $0.141$ & $0.077$ & $0.008$ \\
10\%\, Ada. & $0.362$ & $0.187$ & $0.112$ & $0.009$ \\
\midrule
\,\;\,\;\;\;\;\;\;Sha. & $0.353$ & $0.155$ & $0.072$ & $0.012$ \\
\bottomrule
\end{tabular}
}
\end{table}

\setlength{\tabcolsep}{2.7pt}
\begin{table}[!t]
\caption{Utility in terms of RAQE (with query cardinality scaling for \texttt{COUNT} in the presence of deletion attacks) and MLE Gap under different attacks on Shoppers. The underlying watermarking scheme is \oursys.}
\label{tab:experiments:degrade_shoppers}
\renewcommand{\arraystretch}{0.85}
{\fontfamily{ntxtlf}\fontsize{9pt}{11.5pt}\selectfont
\begin{tabular}{l||c c c c c c c c}
\toprule
\multicolumn{1}{c||}{Attacks} & \makecell{\texttt{COUNT}\\$@1\%$} & \makecell{\texttt{AVG}\\$@1\%$} & \makecell{\texttt{COUNT}\\$@5\%$} & \makecell{\texttt{AVG}\\$@5\%$} & \makecell{\texttt{COUNT}\\\!$@20\%$\!} & \makecell{\texttt{AVG}\\\!$@20\%$\!} & \makecell{MLE\\Gap} \\
\midrule
\,\,w/o\, Att. & $0.322$ & $0.908$ & $0.141$ & $0.419$ & $0.062$ & $0.169$ & $0.009$ \\
\midrule
\;\,5\%\, Gau. & $0.421$ & $0.899$ & $0.211$ & $0.439$ & $0.086$ & $0.174$ & $0.009$ \\
10\%\, Gau. & $0.416$ & $0.900$ & $0.208$ & $0.440$ & $0.084$ & $0.174$ & $0.010$ \\
\midrule
\;\,5\%\, Alt. & $0.348$ & $0.973$ & $0.161$ & $0.437$ & $0.082$ & $0.169$ & $0.014$ \\
10\%\, Alt. & $0.490$ & $1.003$ & $0.237$ & $0.468$ & $0.151$ & $0.172$ & $0.019$ \\
\midrule
\;\,5\%\, Del. & $0.326$ & $0.918$ & $0.142$ & $0.417$ & $0.062$ & $0.169$ & $0.008$ \\
10\%\, Del. & $0.329$ & $0.934$ & $0.144$ & $0.423$ & $0.063$ & $0.173$ & $0.010$ \\
\midrule
\;\,5\%\, Gen. & $0.308$ & $0.934$ & $0.157$ & $0.446$ & $0.075$ & $0.174$ & $0.012$ \\
10\%\, Gen. & $0.363$ & $1.105$ & $0.206$ & $0.611$ & $0.099$ & $0.286$ & $0.015$ \\
\midrule
\;\,5\%\, Ada. & $0.332$ & $0.903$ & $0.152$ & $0.436$ & $0.071$ & $0.200$ & $0.010$ \\
10\%\, Ada. & $0.359$ & $0.924$ & $0.179$ & $0.470$ & $0.090$ & $0.246$ & $0.010$ \\
\midrule
\,\;\,\;\;\;\;\;\;Sha. & $0.497$ & $1.013$ & $0.235$ & $0.524$ & $0.108$ & $0.249$ & $0.021$ \\
\bottomrule
\end{tabular}
}
\end{table}

\subsubsection{Implementation}

\oursys was written in Python 3.10. For clustering, we set $M=256$, and performed K-means clustering on the principal components of tuple latents that accounted for 99\% of the total variance. To estimate the transition matrix $\mathbf{P}$, we set $T=25600$ to ensure that $\eta=1\%$. For watermarked-histogram generation, we used Gurobi~\cite{Gurobi} as optimizer, divided the optimization process into 6 stages, allocated 30 seconds to each stage, and initialized $\tau=0.01$. For w/o WM, TabularMark, and FreqyWM, we synthesized the non-watermarked table according to the original histogram. In our default parameter configuration, we set $N\!=\!1000$, $L\!=\!32$, $\delta_{FPR}\!=\!10^{-3}$ (leading to $\delta_{BE}\!=\!3$), $\delta_{FNR}=10^{-3}$, $I_{per}=0.01, I_{alt}=0.01, I_{del}=0.1$ (except for the Phishing dataset, where $I_{del}$ was set to $0.2$). In addition, we averaged the results on 100 trials.

\subsection{Evaluation on Overall Performance} \label{sec:experiments:robustness_utility}

\setlength{\tabcolsep}{0.82pt}
\begin{table*}[t]
\caption{Comparison of traceability, robustness, and utility between \oursys and baselines. Since Phishing did not have numerical attributes, we did not launch perturbation attacks or measure RAQE for \texttt{AVG} queries on this dataset (denoted by N/A). Moreover, FreqyWM failed to encode 32-bit watermarks on Phishing (denoted by $\times$).}
\label{tab:experiments:quality_detection}
{\fontfamily{ntxtlf}\fontsize{8.7pt}{11pt}\selectfont
\begin{tabular}{c|l||c||c|c|c|c|c|c|c|c|c||c|c|c|c|c|c|c|c|c}
\toprule
\multicolumn{1}{c|}{\multirow{4}{*}{Schemes}} & \multicolumn{1}{c||}{\multirow{4}{*}{Datasets}} & \multirow{4}{*}{\makecell{Trace-\\ability\\(Acc.\!${\uparrow}$)}} & \multicolumn{9}{c||}{Robustness (Acc.${\uparrow}$)} & \multicolumn{9}{c}{Utility} \\
\cline{4-21}
& & & \multirow{3}{*}{\makecell{$5\%$\\Gau.}} & \multirow{3}{*}{\makecell{$5\%$\\Uni.}} & \multirow{3}{*}{\makecell{$5\%$\\Lap.}} & \multirow{3}{*}{\makecell{$5\%$\\\,Alt.\,}} & \multirow{3}{*}{\makecell{$5\%$\\\,Del.\,}} & \multirow{3}{*}{\makecell{$5\%$\\\,Ins.\,}} & \multirow{3}{*}{\makecell{$5\%$\\Gen.}} & \multirow{3}{*}{\makecell{$5\%$\\Ada.}} & \multirow{3}{*}{\makecell{Sha.}} & \multirow{3}{*}{\makecell{\,Marg.\,\\Gap\,${\downarrow}$}} & \multirow{3}{*}{\makecell{\,Corr.\,\\Gap\,${\downarrow}$}} & \multirow{3}{*}{\makecell{\,\,MLE\,\,\\Gap\,${\downarrow}$}} & \multicolumn{6}{c}{95-th RAQE at Different Selectivities\,$\downarrow$} \\
\cline{16-21}
& & & & & & & & & & & & & & & \multicolumn{2}{c|}{\multirow{1}{*}{$1\%$ (Low)}} & \multicolumn{2}{c|}{\multirow{1}{*}{$5\%$\! (Medium)}} & \multicolumn{2}{c}{\multirow{1}{*}{$20\%$ (High)}} \\
\cline{16-21}
& & & & & & & & & & & & & & & \texttt{COUNT} & \texttt{\,\;AVG\;\,} & \texttt{COUNT} & \texttt{\,\;AVG\;\,} & \texttt{COUNT} & \texttt{\,\;AVG\;\,} \\
\midrule
\multirow{4}{*}{\makecell{w/o\\WM}} & Beijing & N/A & N/A & N/A & N/A & N/A & N/A & N/A & N/A & N/A & N/A & $0.022$ & $0.027$ & $0.109$ & $0.231$ & $0.722$ & $0.172$ & $0.341$ & $0.060$ & $0.308$ \\
& Electric & N/A & N/A & N/A & N/A & N/A & N/A & N/A & N/A & N/A & N/A & $0.013$ & $0.007$ & $0.017$ & $0.260$ & $0.119$ & $0.118$ & $0.050$ & $0.054$ & $0.022$ \\
& Phishing & N/A & N/A & N/A & N/A & N/A & N/A & N/A & N/A & N/A & N/A & $0.005$ & $0.009$ & $0.008$ & $0.269$ & N/A & $0.115$ & N/A & $0.057$ & N/A \\
& Shoppers & N/A & N/A & N/A & N/A & N/A & N/A & N/A & N/A & N/A & N/A & $0.020$ & $0.024$ & $0.008$ & $0.320$ & $0.900$ & $0.140$ & $0.418$ & $0.062$ & $0.170$ \\
\midrule
\midrule
\multirow{4}{*}{\makecell{Tabular-\\Mark}} & Beijing & $\textbf{1.00}$ & $\textbf{1.00}$ & $\textbf{1.00}$ & $\textbf{1.00}$ & $\textbf{1.00}$ & $\textbf{1.00}$ & $\textbf{1.00}$ & $0.99$ & $\textbf{1.00}$ & $0.00$ & $\textbf{0.022}$ & $0.030$ & $0.114$ & $0.234$ & $\textbf{0.724}$ & $0.181$ & $0.341$ & $0.069$ & $0.307$ \\
& Electric & $\textbf{1.00}$ & $\textbf{1.00}$ & $\textbf{1.00}$ & $\textbf{1.00}$ & $\textbf{1.00}$ & $\textbf{1.00}$ & $\textbf{1.00}$ & $0.95$ & $\textbf{1.00}$ & $0.00$ & $\textbf{0.013}$ & $0.008$ & $\textbf{0.017}$ & $\textbf{0.261}$ & $\textbf{0.118}$ & $\textbf{0.118}$ & $\textbf{0.050}$ & $\textbf{0.054}$ & $\textbf{0.023}$ \\
& Phishing & $\textbf{1.00}$ & N/A & N/A & N/A & $0.94$ & $\textbf{1.00}$ & $\textbf{1.00}$ & $\textbf{1.00}$ & $\textbf{1.00}$ & $0.00$ & $0.006$ & $0.011$ & $\textbf{0.008}$ & $0.299$ & N/A & $0.152$ & N/A & $0.102$ & N/A \\
& Shoppers & $\textbf{1.00}$ & $\textbf{1.00}$ & $\textbf{1.00}$ & $\textbf{1.00}$ & $\textbf{1.00}$ & $\textbf{1.00}$ & $\textbf{1.00}$ & $0.97$ & $\textbf{1.00}$ & $0.00$ & $0.022$ & $\textbf{0.023}$ & $\textbf{0.009}$ & $0.327$ & $0.917$ & $0.144$ & $0.433$ & $0.076$ & $0.171$ \\
\midrule
\multirow{4}{*}{\makecell{Freqy-\\WM}} & Beijing & $\textbf{1.00}$ & $0.00$ & $0.00$ & $0.00$ & $0.00$ & $0.00$ & $0.00$ & $0.00$ & $0.00$ & $0.00$ & $\textbf{0.022}$ & $\textbf{0.027}$ & $0.109$ & $\textbf{0.231}$ & $0.728$ & $0.173$ & $\textbf{0.339}$ & $0.060$ & $\textbf{0.305}$ \\
& Electric & $\textbf{1.00}$ & $0.00$ & $0.00$ & $0.00$ & $0.00$ & $0.00$ & $0.00$ & $0.00$ & $0.00$ & $0.00$ & $\textbf{0.013}$ & $\textbf{0.007}$ & $\textbf{0.017}$ & $\textbf{0.261}$ & $0.119$ & $0.119$ & $\textbf{0.050}$ & $0.055$ & $\textbf{0.023}$ \\
& Phishing & $\times$ & N/A & N/A & N/A & $\times$ & $\times$ & $\times$ & $\times$ & $\times$ & $\times$ & $\times$ & $\times$ & $\times$ & $\times$ & N/A & $\times$ & N/A & $\times$ & N/A \\
& Shoppers & $\textbf{1.00}$ & $0.00$ & $0.00$ & $0.00$ & $0.00$ & $0.00$ & $0.00$ & $0.00$ & $0.00$ & $0.00$ & $\textbf{0.020}$ & $0.024$ & $\textbf{0.009}$ & $\textbf{0.322}$ & $\textbf{0.905}$ & $\textbf{0.139}$ & $0.420$ & $\textbf{0.062}$ & $0.175$ \\
\midrule
\multirow{4}{*}{TabWak} & Beijing & $0.00$ & $0.00$ & $0.00$ & $0.00$ & $0.00$ & $0.00$ & $0.00$ & $0.00$ & $0.00$ & $0.00$ & $0.074$ & $0.112$ & $0.359$ & $0.993$ & $1.573$ & $0.741$ & $1.753$ & $0.305$ & $1.627$ \\
& Electric & $0.00$ & $0.00$ & $0.00$ & $0.00$ & $0.00$ & $0.00$ & $0.00$ & $0.00$ & $0.00$ & $0.00$ & $0.077$ & $0.042$ & $0.072$ & $1.790$ & $0.250$ & $0.958$ & $0.178$ & $0.410$ & $0.147$ \\
& Phishing & $0.00$ & N/A & N/A & N/A & $0.00$ & $0.00$ & $0.00$ & $0.00$ & $0.00$ & $0.00$ & $0.051$ & $0.081$ & $0.018$ & $0.919$ & N/A & $0.706$ & N/A & $0.457$ & N/A \\
& Shoppers & $0.00$ & $0.00$ & $0.00$ & $0.00$ & $0.00$ & $0.00$ & $0.00$ & $0.00$ & $0.00$ & $0.00$ & $0.080$ & $0.098$ & $0.037$ & $1.451$ & $7.070$ & $0.921$ & $4.576$ & $0.424$ & $3.231$ \\
\midrule
\multirow{4}{*}{\makecell{\texttt{Table-}\\\texttt{Mark}}} & Beijing & $\textbf{1.00}$ & $\textbf{1.00}$ & $\textbf{1.00}$ & $\textbf{1.00}$ & $\textbf{1.00}$ & $\textbf{1.00}$ & $\textbf{1.00}$ & $\textbf{1.00}$ & $0.99$ & $0.00$ & $\textbf{0.022}$ & $\textbf{0.027}$ & $\textbf{0.108}$ & $0.232$ & $0.727$ & $\textbf{0.172}$ & $0.342$ & $\textbf{0.059}$ & $0.308$ \\
& Electric & $\textbf{1.00}$ & $\textbf{1.00}$ & $\textbf{1.00}$ & $\textbf{1.00}$ & $\textbf{1.00}$ & $\textbf{1.00}$ & $\textbf{1.00}$ & $\textbf{1.00}$ & $0.99$ & $0.00$ & $\textbf{0.013}$ & $0.008$ & $\textbf{0.017}$ & $\textbf{0.261}$ & $\textbf{0.118}$ & $\textbf{0.118}$ & $\textbf{0.050}$ & $0.055$ & $\textbf{0.023}$ \\
& Phishing & $\textbf{1.00}$ & N/A & N/A & N/A & $\textbf{1.00}$ & $\textbf{1.00}$ & $\textbf{1.00}$ & $\textbf{1.00}$ & $\textbf{1.00}$ & $0.00$ & $\textbf{0.005}$ & $\textbf{0.009}$ & $\textbf{0.008}$ & $\textbf{0.276}$ & N/A & $\textbf{0.120}$ & N/A & $\textbf{0.060}$ & N/A \\
& Shoppers & $\textbf{1.00}$ & $\textbf{1.00}$ & $\textbf{1.00}$ & $\textbf{1.00}$ & $\textbf{1.00}$ & $\textbf{1.00}$ & $\textbf{1.00}$ & $\textbf{1.00}$ & $\textbf{1.00}$ & $0.00$ & $\textbf{0.020}$ & $0.024$ & $\textbf{0.009}$ & $\textbf{0.322}$ & $0.908$ & $0.141$ & $\textbf{0.419}$ & $\textbf{0.062}$  & $\textbf{0.169}$ \\
\bottomrule
\end{tabular}
}
\end{table*}

\begin{figure}[!t]
  \centering
  \includegraphics[width=0.9\linewidth]{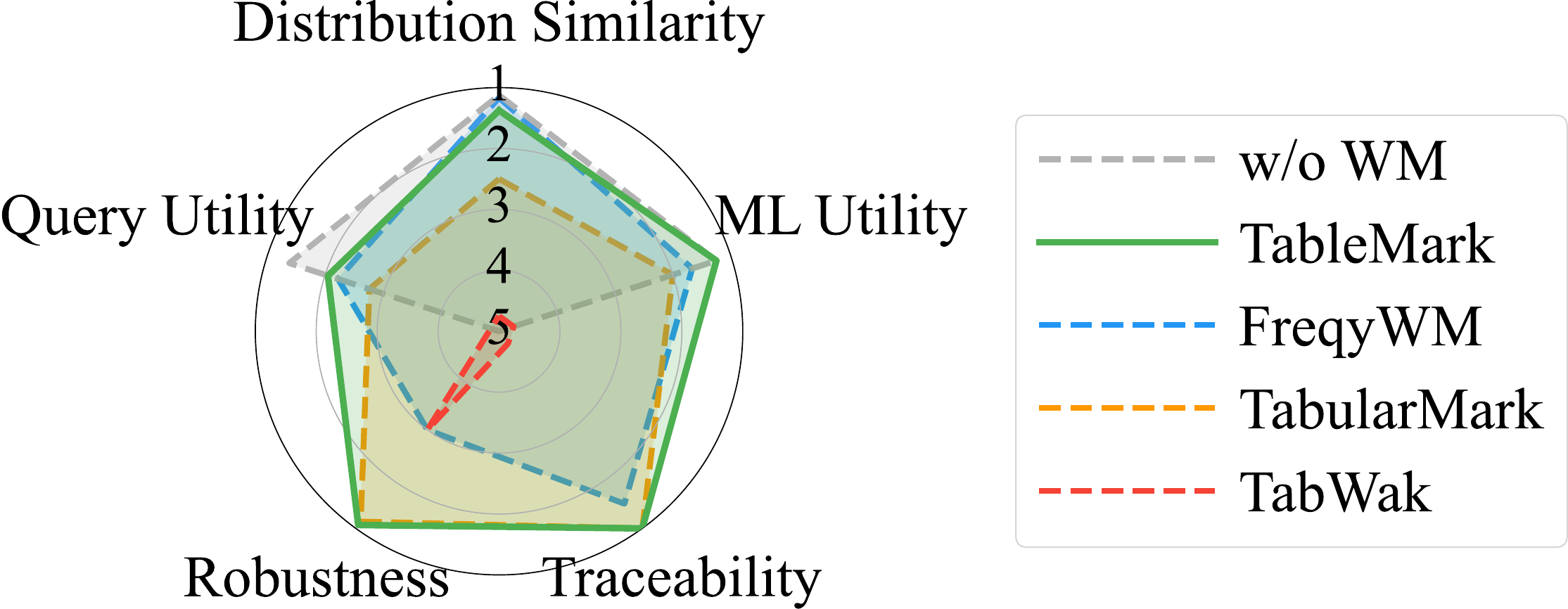}
  \caption{Average ranking comparison between \oursys and baselines across distribution similarity between the synthetic and original tables, machine learning utility, query utility, traceability, and robustness. Outer regions indicate higher ranks and better performance.}
  \label{fig:experiments:rank}
\end{figure}

To answer \textbf{RQ1}, we investigated the traceability, robustness, and utility of \oursys and baselines. Figure~\ref{fig:experiments:rank} and Table~\ref{tab:experiments:quality_detection} show the evaluation results.

For traceability, we made the following observations. Firstly, TabWak had the lowest accuracy (almost 0), while TabularMark, \!FreqyWM, \!and \oursys achieved the highest accuracy (almost 1). The reason TabWak had low traceability was that the process of recovering noise vectors during watermark decoding was error-prone, due to limitations of DDIM. As a result, when encoding multiple bits (32 bits), TabWak could not accurately extract the noise pattern of each bit. For TabularMark and FreqyWM, they did not involve watermark decoding errors and could achieve perfect accuracy without attacks. For \oursys, the constraint generation mechanism calibrated watermark decoding errors (i.e., cluster misassignments caused by the reconstruction error of the tuple embedding model, see Section~\ref{sec:constraint:error_attack}). Secondly, FreqyWM failed to encode 32-bit watermarks on Phishing (denoted by $\times$ in Table~\ref{tab:experiments:quality_detection}), while \oursys achieved high accuracy on this dataset. This was because FreqyWM maintained the ranking of partition sizes between the watermarked and non-watermarked synthetic tables for utility, so there was limited room to modify the non-watermarked synthetic table for watermark encoding. Moreover, Phishing was a small dataset with fewer than 10000 tuples in the original table, further reducing the modification room. In contrast, \oursys did not involve restrictive utility constraints, but instead treated utility as the optimization goal (i.e., minimizing the difference between the watermarked and original histograms), so \oursys could accommodate a large number of bits.

For robustness, we made the following observations. Firstly, TabWak and FreqyWM had the lowest traceability accuracy (almost 0) in the presence of attacks. For TabWak, it could not effectively recover noise vectors during watermark decoding, as discussed above. For FreqyWM, its congruence-relation patterns were sensitive to partition-size fluctuations caused by attacks. Secondly, both TabularMark and \oursys achieved high accuracy (almost 1) under attacks. For TabularMark, it injected large scale noise (uniform noise with standard deviations set to 20\% of attributes' standard deviations for numerical attributes), which overwhelmed the noise introduced by perturbation attacks. In addition, TabularMark injected the noise into a moderate proportion of cells (15\% cells within an attribute), while alteration, deletion, and insertion attacks were targeted at a small proportion of cells or tuples (5\% cells or tuples), so TabularMark could resist these attacks. For \oursys, the constraint generation mechanism ensured enough gaps between cluster sizes of partial orders, so these partial orders were more resilient.

For utility, we obtained the following observations. Firstly, TabWak had the lowest performance (e.g., a degradation in Marg. Gap by 0.046-0.064 relative to w/o WM). This was because to support multi-bit watermarks, TabWak encoded complex patterns into noise vectors that corresponded to individual tuples, resulting in significant distortions to each tuple. Secondly, TabularMark exhibited lower performance, especially in terms of RAQE on high-dimensional datasets (e.g., an increase of 0.030-0.045 on Phishing and around 0.010 on Shoppers compared to w/o WM). This was because to ensure robustness, TabularMark performed substantial modifications (large-scale noise for numerical attributes and value substitutions for categorical attributes) to a moderate proportion of cells (15\% cells within an attribute), which significantly compromised tuple integrity and lowered the accuracy of high-dimensional queries that involved multiple predicates applied to individual tuples. Notably, TabularMark exhibited lower utility when watermarks were encoded into categorical attributes (on Phishing) than into numerical attributes (on other datasets), because categorical attributes had small and discrete domains, so modifications to them often introduced greater distortions. This suggested that TabularMark was less suitable for categorical datasets. Finally, FreqyWM and \oursys achieved the highest utility comparable to w/o WM (with a degradation of less than 0.002 on almost all metrics). For FreqyWM, it imposed the restrictive ranking constraint on partition sizes. For \oursys, it ensured tuple integrity, and preserved high-quality tuple-frequency distributions by designing effective optimization mechanisms.

In summary, \oursys significantly outperforms the state-of-the-art baselines, demonstrating effective traceability, strong robustness, and high utility across various types of datasets.

\subsection{Evaluation on Constraint Generator} \label{sec:experiments:constraint}

\begin{figure}[!t]
  \centering
  \includegraphics[width=1\linewidth]{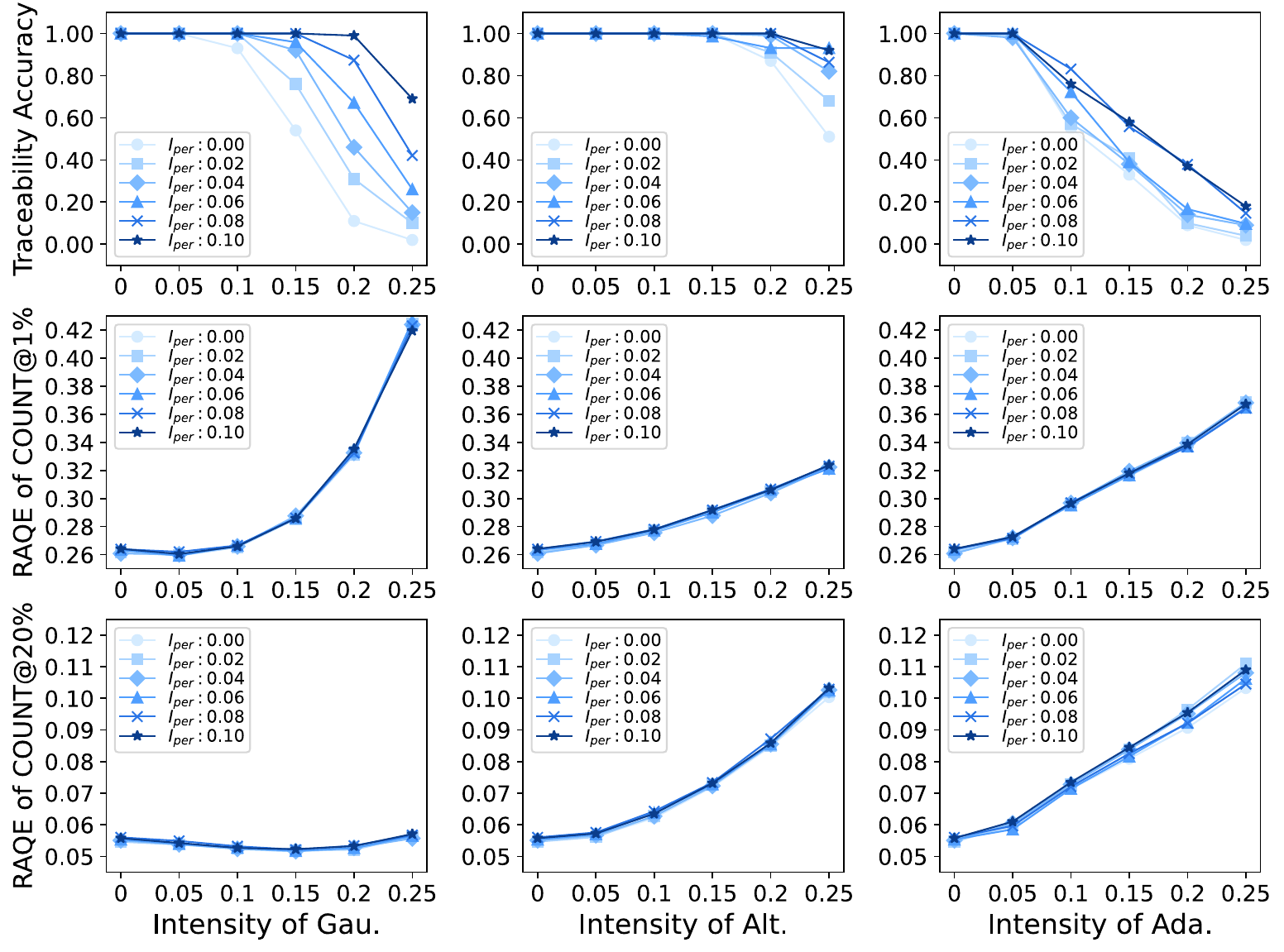}
  \caption{Comparison of traceability accuracy under Gaussian perturbation, alteration, and adaptive tuple-deletion attacks, and RAQE of \texttt{COUNT} queries (with cardinality scaling under adaptive tuple-deletion attacks) at selectivities 1\% and 20\%, over different $I_{per}$ values.}
  \label{fig:experiments:gauss}
  \vspace{-.5em}
\end{figure}

\begin{figure}[!t]
  \centering
  \includegraphics[width=1\linewidth]{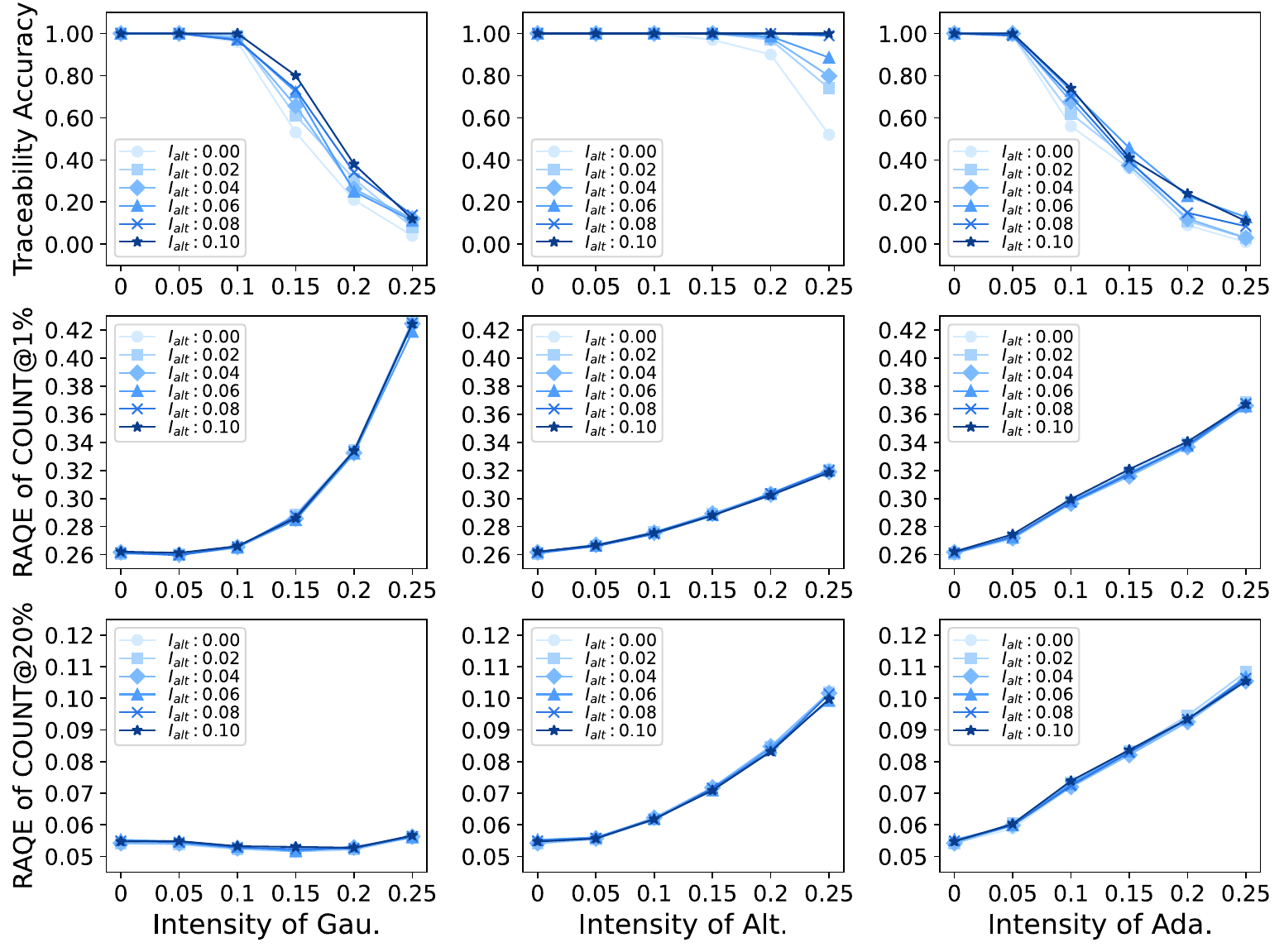}
  \caption{Comparison of traceability accuracy under Gaussian perturbation, alteration, and adaptive tuple-deletion attacks, and RAQE of \texttt{COUNT} queries (with cardinality scaling under adaptive tuple-deletion attacks) at selectivities 1\% and 20\%, over different $I_{alt}$ values.}
  \label{fig:experiments:alter}
\end{figure}

\begin{figure}[!t]
  \centering
  \includegraphics[width=1\linewidth]{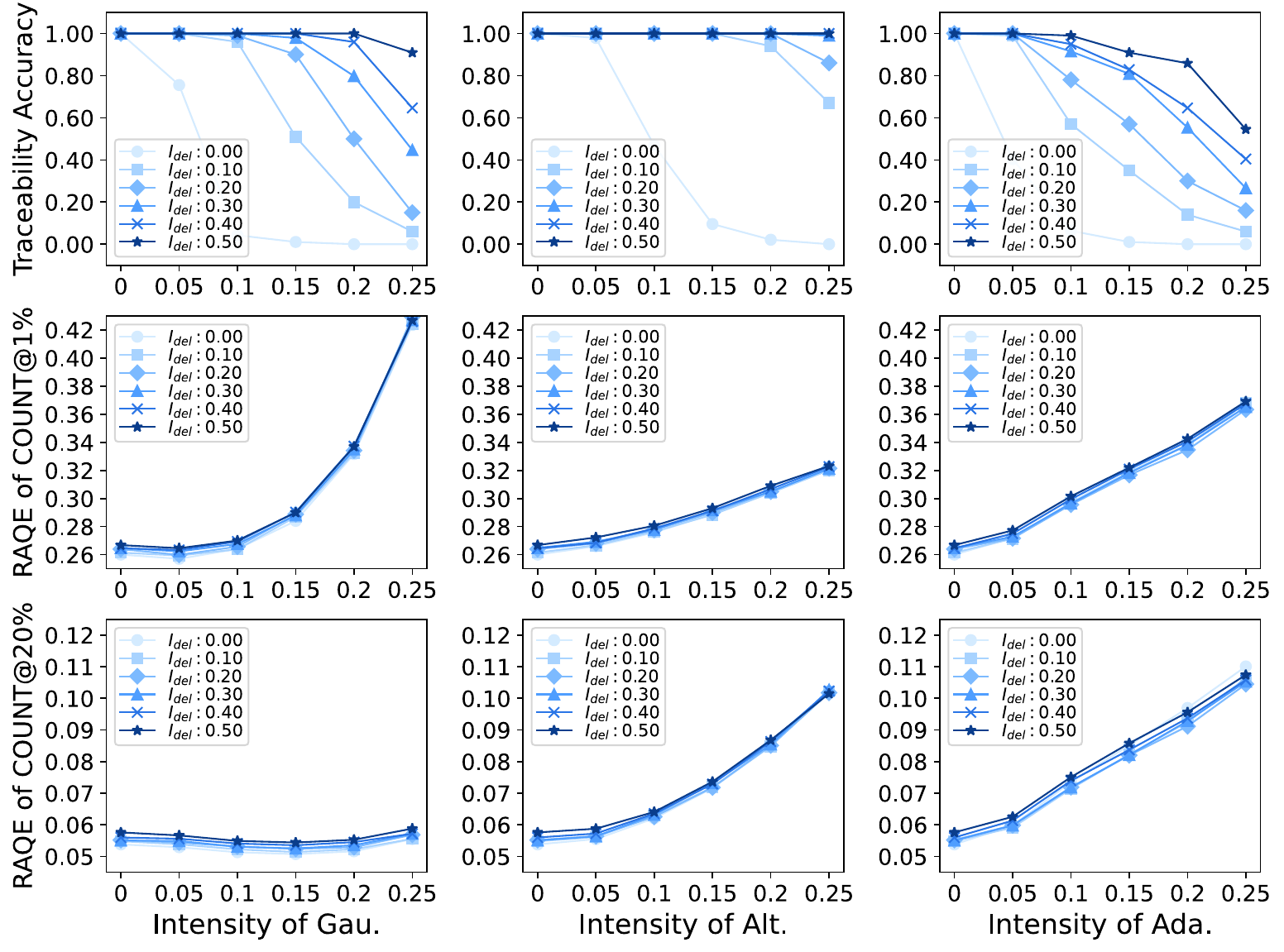}
  \caption{Comparison of traceability accuracy under Gaussian perturbation, alteration, and adaptive tuple-deletion attacks, and RAQE of \texttt{COUNT} queries (with cardinality scaling under adaptive tuple-deletion attacks) at selectivities 1\% and 20\%, over different $I_{del}$ values.}
  \label{fig:experiments:deletion_rate}
\end{figure}

\begin{figure}[!t]
  \centering
  \includegraphics[width=1\linewidth]{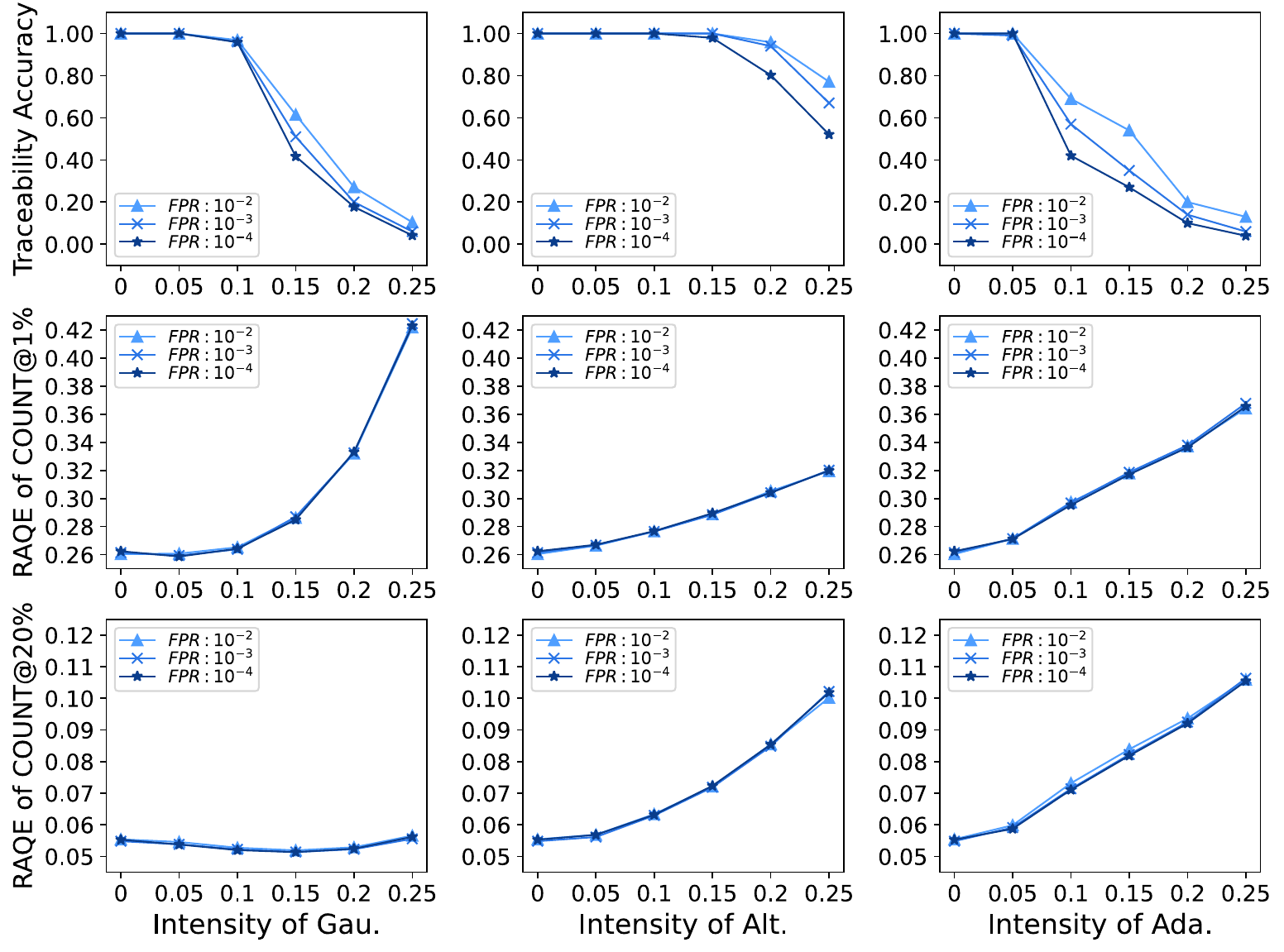}
  \caption{Comparison of traceability accuracy under Gaussian perturbation, alteration, and adaptive tuple-deletion attacks, and RAQE of \texttt{COUNT} queries (with cardinality scaling under adaptive tuple-deletion attacks) at selectivities 1\% and 20\%, over different $\delta_{FPR}$ values.}
  \label{fig:experiments:fpr}
\end{figure}

\begin{figure}[!t]
  \centering
  \includegraphics[width=1\linewidth]{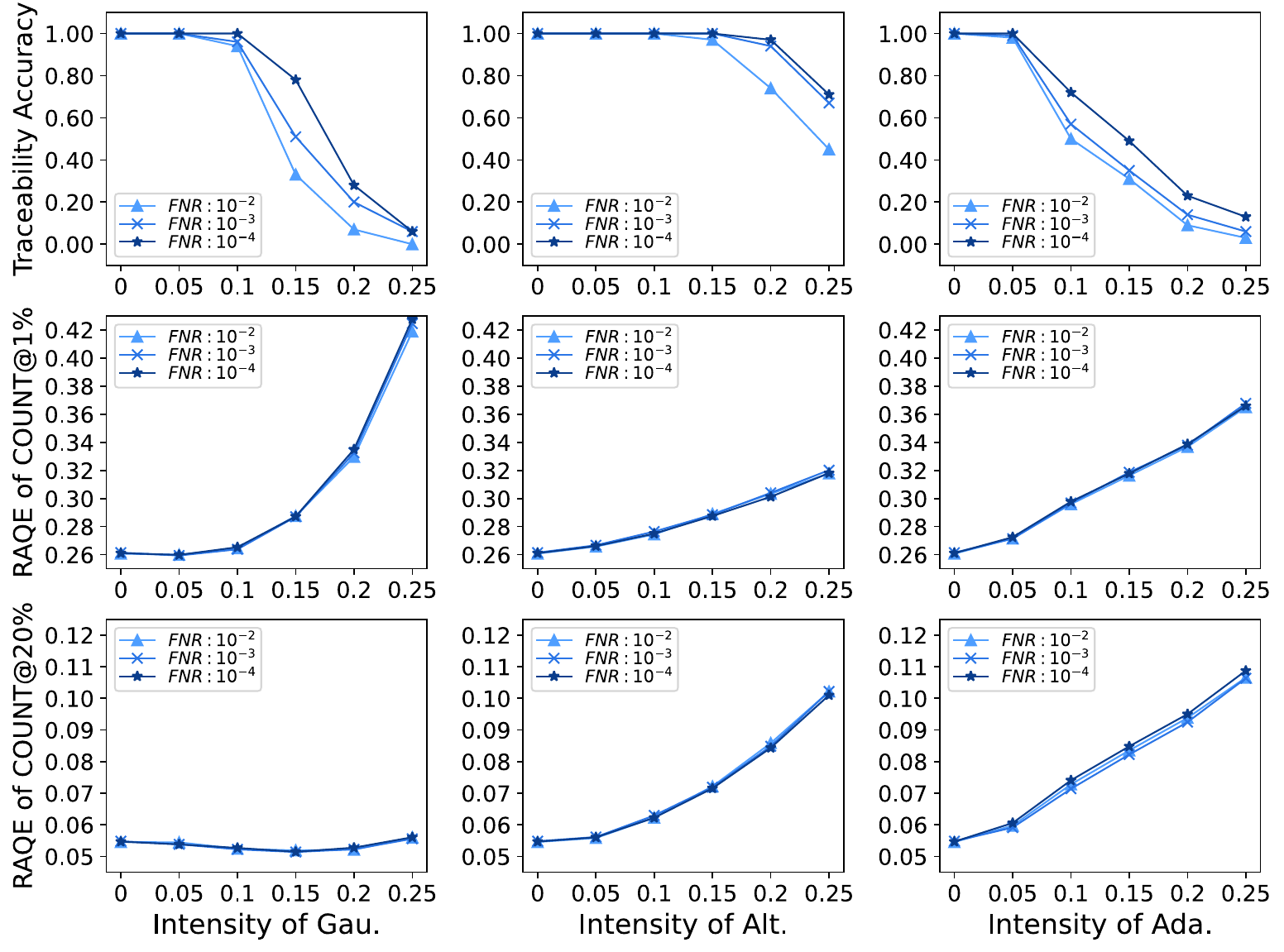}
  \caption{Comparison of traceability accuracy under Gaussian perturbation, alteration, and adaptive tuple-deletion attacks, and RAQE of \texttt{COUNT} queries (with cardinality scaling under adaptive tuple-deletion attacks) at selectivities 1\% and 20\%, over different $\delta_{FNR}$ values.}
  \label{fig:experiments:fnr}
\end{figure}

To answer \textbf{RQ2}, we investigated the robustness and utility of \oursys over different values of the robustness parameters. For robustness, we measured the traceability accuracy under three typical types of attacks: cell-level Gaussian perturbation and alteration attacks that modified certain cells, and tuple-level adaptive deletion attacks that removed certain tuples. For utility, we measured the RAQE of \texttt{COUNT} queries (with cardinality scaling under deletion attacks) at selectivities 1\% and 20\%. We conducted the experiments on the Electric dataset and the results are shown in Figure~\ref{fig:experiments:gauss} (w.r.t. $I_{per}$), Figure~\ref{fig:experiments:alter} (w.r.t. $I_{alt}$), Figure~\ref{fig:experiments:deletion_rate} (w.r.t. $I_{del}$), Figure~\ref{fig:experiments:fpr} (w.r.t. $\delta_{FPR}$), and Figure~\ref{fig:experiments:fnr} (w.r.t. $\delta_{FNR}$).

For $I_{per}$, $I_{alt}$, and $I_{del}$, we made the following observations. Firstly, as $I_{per}$ increased, there was an obvious increase in the traceability accuracy under perturbation attacks, and a slight increase under alteration and adaptive tuple-deletion attacks. Similarly, as $I_{alt}$ increased, the accuracy under alteration attacks exhibited a significant increase, while the accuracy under perturbation and adaptive tuple-deletion attacks showed a slight increase. Moreover, as $I_{del}$ increased, the accuracy under all three classes of attacks experienced a substantial increase. Secondly, the increase in $I_{per}$ or $I_{alt}$ caused a slight utility loss, while the increase in $I_{del}$ resulted in a more obvious utility loss (in Figures~\ref{fig:experiments:gauss} and~\ref{fig:experiments:alter}, the curves for different values of $I_{per}$ or $I_{alt}$ almost overlapped in the second and third rows, while in Figure~\ref{fig:experiments:deletion_rate}, the curves with larger values of $I_{del}$ lay clearly above those with smaller ones). Thirdly, compared to the utility loss caused by increasing $I_{per}$, $I_{alt}$, or $I_{del}$, the utility loss incurred by attacks was substantially more severe, implying that \oursys could significantly enhance robustness at a marginal utility cost.

For $\delta_{FPR}$ and $\delta_{FNR}$, we made the following observations. Firstly, traceability accuracy decreased as $\delta_{FPR}$ decreased, because a smaller $\delta_{FPR}$ meant a reduced bit error tolerance (see Equation~\ref{eq:FPR2}). Secondly, traceability accuracy slightly increased as $\delta_{FNR}$ decreased, because a smaller $\delta_{FNR}$ led to a smaller $\delta_{BER}$ (see Equation~\ref{eq:FNR2}) and thus larger gaps between cluster sizes of partial orders (see Equation~\ref{eq:xlr}), which make partial orders more resilient. Thirdly, changes in $\delta_{FPR}$ or $\delta_{FNR}$ each led to minimal utility loss (RAQE values were almost identical over different values of $\delta_{FPR}$ or $\delta_{FNR}$ in Figures~\ref{fig:experiments:fpr} and~\ref{fig:experiments:fnr}). Note that different combinations of $I_{per}$, $I_{alt}$, and $I_{del}$ can already lead to various trade-offs between robustness and utility (discussed above), we recommend that the data owner configure $\delta_{FPR}$ according to his/her desired level of false positives and set $\delta_{FNR}$ to $0.001$ as in our default configuration.

In summary, \oursys allows the data owner to effectively trade off robustness and utility, where the data owner configures $\delta_{FPR}$ based on his/her expected level of false positives, sets $\delta_{FNR}$ to $0.001$, sets $I_{per}$ and $I_{alt}$ to small values to ensure basic robustness against perturbation and alteration attacks, and finally chooses a value for $I_{del}$ to achieve a desired level of robustness against various attacks at the cost of minimal utility.

\subsection{Evaluation on Optimization Mechanisms} \label{sec:experiments:optimizer}

\begin{figure}[!t]
  \centering
  \includegraphics[width=0.95\linewidth]{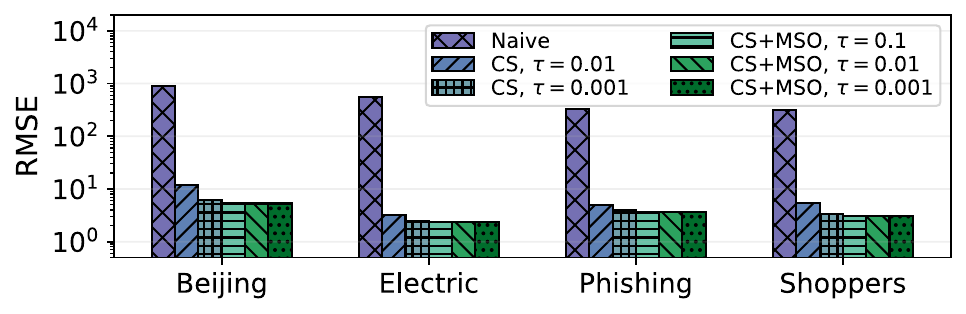}
  \caption{RMSE difference between the watermarked and original histograms (lower values indicate higher utility) over various $\tau$ initializations, across different combinations of optimization mechanisms: (1) without optimization mechanisms (Naive), (2) constraint simplification mechanism (CS), (3) constraint simplification and multi-stage optimization mechanisms (CS+MSO).}
  \label{fig:experiments:opt}
\end{figure}

To answer \textbf{RQ3}, we investigated the RMSE difference between the watermarked and original histograms under different combinations of optimization mechanisms over various initial values of $\tau$. The results are shown in Figure~$\ref{fig:experiments:opt}$. We made the following observations. Firstly, compared to the naive method, the constraint simplification mechanism significantly reduced RMSE (by almost two orders of magnitude), because this mechanism led to a more tractable optimization problem. Secondly, with the multi-stage optimization mechanism, RMSE dropped further (to about 0.4-0.8x), because $\tau$ was gradually adjusted to appropriate values as optimization progressed. Thirdly, with the multi-stage optimization mechanism, although $\tau$ was initialized to different values (0.001, 0.01, and 0.1), the resulting RMSE values were close, because $\tau$ was adaptively adjusted during optimization regardless of its initial values.

In summary, the constraint simplification and multi-stage optimization mechanisms significantly improve the data utility of the watermarked table and are not sensitive to the initial value of $\tau$.

\subsection{Analysis on $M$, $N$, and $L$} \label{sec:experiments:mnl}

\begin{figure}[!t]
  \centering
  \includegraphics[width=1\linewidth]{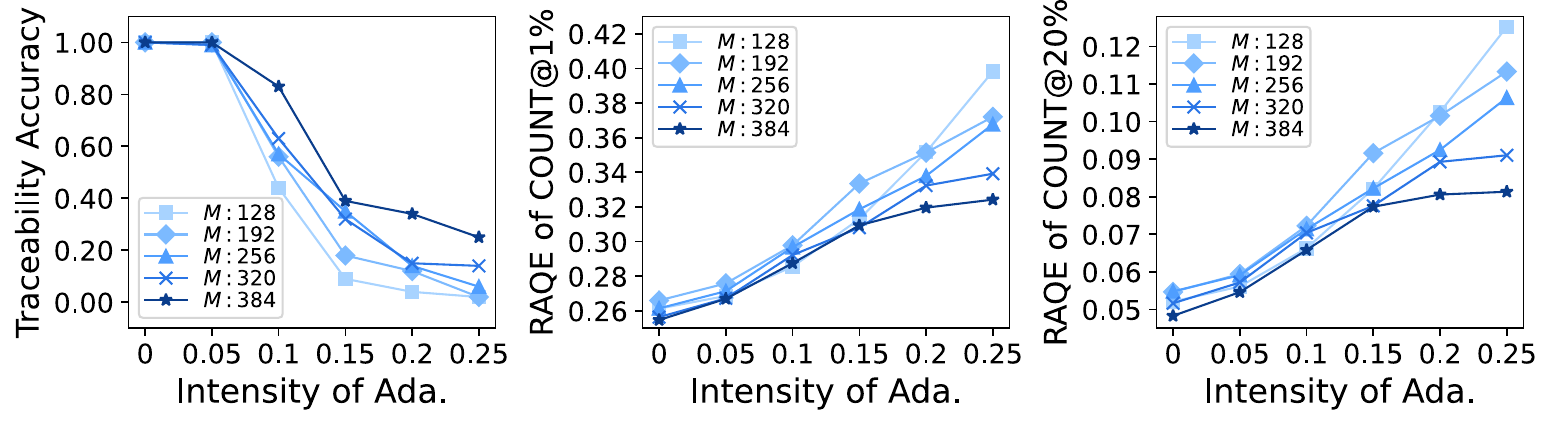}
  \caption{Comparison of traceability accuracy under adaptive tuple-deletion attacks and RAQE of \texttt{COUNT} queries (with cardinality scaling) at selectivities 1\% and 20\%, over different $M$ values.}
  \label{fig:experiments:num_classes}
\end{figure}

\begin{figure}[!t]
  \centering
  \includegraphics[width=1\linewidth]{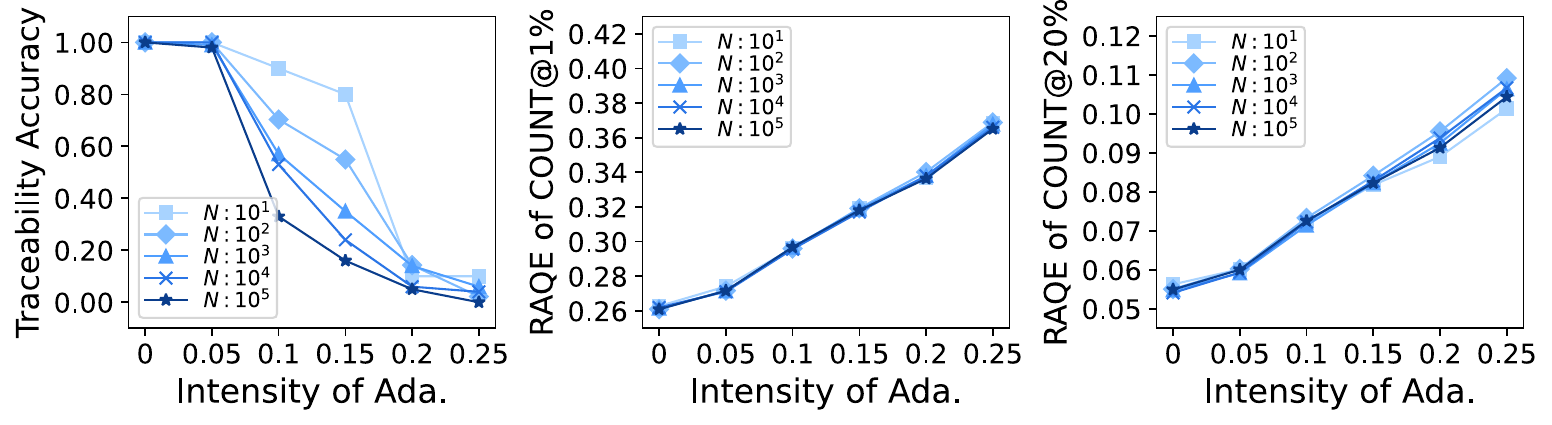}
  \caption{Comparison of traceability accuracy under adaptive tuple-deletion attacks and RAQE of \texttt{COUNT} queries (with cardinality scaling) at selectivities 1\% and 20\%, over different $N$ values.}
  \label{fig:experiments:num_users}
\end{figure}

\begin{figure}[!t]
  \centering
  \includegraphics[width=1\linewidth]{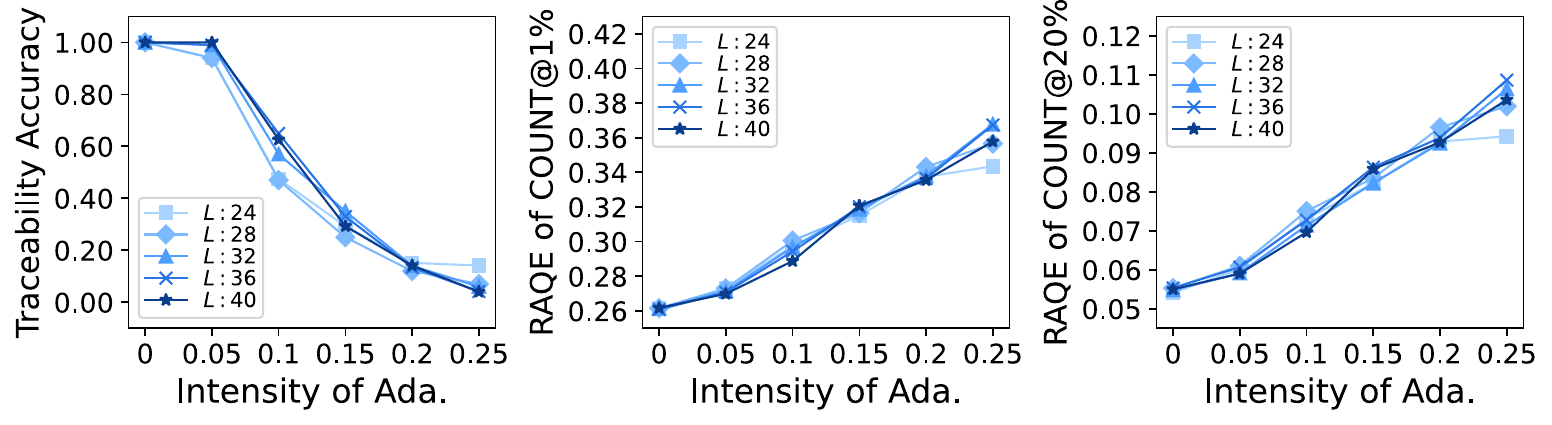}
  \caption{Comparison of traceability accuracy under adaptive tuple-deletion attacks and RAQE of \texttt{COUNT} queries (with cardinality scaling) at selectivities 1\% and 20\%, over different $L$ values.}
  \label{fig:experiments:num_bits}
\end{figure}

To answer \textbf{RQ4}, we investigated the robustness and utility of \oursys over different values of $M$, $N$, and $L$. For robustness, we measured the traceability accuracy under adaptive tuple-deletion attacks. For utility, we measured the RAQE of \texttt{COUNT} queries (with cardinality scaling) at selectivities 1\% and 20\%. The experiments were conducted on the Electric dataset and the results are shown in Figure~\ref{fig:experiments:num_classes} (w.r.t. $M$), Figure~\ref{fig:experiments:num_users} (w.r.t. $N$), and Figure~\ref{fig:experiments:num_bits} (w.r.t. $L$).

\subsubsection{Analysis on $M$}

We obtained the following observations from Figure~\ref{fig:experiments:num_classes}. Firstly, traceability accuracy increased with $M$, because a larger $M$ increased the clustering randomness and reduced the proportion of tuples within the $2L$ template clusters, making adaptive deletion operations less likely to hit the template clusters. This observation indicated that a larger $M$ led to better robustness. Secondly, RAQE decreased as $M$ increased, implying that a larger $M$ also led to better utility. This was because the original table was partitioned into finer-grained clusters, making the distribution of each cluster simpler and thus easier for the conditional latent diffusion model to learn. Consequently, the synthetic table better resembled the original table in terms of intra-cluster distributions.

\setlength{\tabcolsep}{1.9pt}
\begin{table}[t]
\centering
\caption{Privacy risks in terms of the AUROC score and TPR@1\%FPR score over different $M$ values (higher AUROC scores above 0.5 and higher TPR@1\%FPR scores above 0.01 indicate higher privacy risks).}
\label{tab:experiments:mia}
\renewcommand{\arraystretch}{0.9}
{\fontfamily{ntxtlf}\fontsize{9pt}{11.5pt}\selectfont
\begin{tabular}{l l||c c c c}
\toprule
Datasets & Metrics & $M\!=\!64$ & $M\!=\!128$ & $M\!=\!256$ & $M\!=\!512$ \\
\midrule
\multirow{2}{*}{Beijing} & AUROC & 0.536 & 0.547 & 0.555 & 0.565 \\
& TPR$@1\%$FPR & 0.017 & 0.020 & 0.022 & 0.025 \\
\midrule
\multirow{2}{*}{Electric} & AUROC & 0.530 & 0.540 & 0.549 & 0.567 \\
& TPR$@1\%$FPR & 0.013 & 0.015 & 0.015 & 0.020 \\
\midrule
\multirow{2}{*}{Phishing} & AUROC & 0.510 & 0.522 & 0.538 & 0.558 \\
& TPR$@1\%$FPR & 0.000 & 0.000 & 0.000 & 0.000 \\
\midrule
\multirow{2}{*}{Shoppers} & AUROC & 0.512 & 0.515 & 0.520 & 0.528 \\
& TPR$@1\%$FPR & 0.014 & 0.015 & 0.016 & 0.016 \\
\bottomrule
\end{tabular}
}
\end{table}

Although a larger $M$ resulted in better robustness and utility, we argued that a larger $M$ also led to higher privacy risks. This was because as $M$ increased, the number of tuples within each cluster decreased, so the conditional latent diffusion model was more likely to ``memorize'' tuples in each cluster rather than learn the intra-cluster distributions. To validate this claim, following~\cite{MIA}, we measured empirical privacy risks of the synthetic table (specifically, non-watermarked synthetic table because watermarking often did not affect privacy risks) using membership inference attacks, where the attacker was in possession of the synthetic table and tried to infer whether certain tuples (called target tuples) belonged to the original table, by computing the distance between target tuples and tuples from the synthetic table. Specifically, we mixed tuples from the original table and the test set to form target tuples, normalized all attributes, and calculated the Euclidean distance between all target tuples and tuples from the synthetic table. Given a distance threshold, a target tuple was predicted to be in the original table if there existed at least one tuple in the synthetic table with a distance smaller than this threshold. Since the distance threshold could vary, we reported the AUROC score and the TPR@1\%FPR score, as shown in Table~\ref{tab:experiments:mia}. It could be seen that both scores increased with $M$, suggesting that a larger $M$ resulted in higher privacy risks.

In summary, a larger $M$ leads to better robustness and utility, but also results in higher privacy risks. Therefore, we recommend that the data owner choose a large number for $M$ as long as the empirical privacy risks are acceptable.

\setlength{\tabcolsep}{2.7pt}
\begin{table*}[!t]
\caption{Runtime efficiency of \oursys in terms of seconds across different datasets.}
\centering
\label{tab:experiments:time}
{\fontsize{9pt}{11.5pt}\selectfont
\begin{tabular}{c|l||r r r r}
\toprule
\multicolumn{2}{c||}{Stages} & Beijing & Electric & Phishing & Shoppers \\
\midrule
\multirow{5}{*}{\makecell{Watermark-channel\\construction (offline)}} & \makecell[l]{Tuple embedding model training} & 2560.28 & 997.99 & 2076.62 & 1366.85 \\
& Clustering & 302.14 & 56.73 & 95.70 & 55.56 \\
& Conditional latent diffusion model training & 1408.31 & 756.17 & 737.17 & 762.27 \\
& Cluster-pair-based watermark-template construction & 0.01 & 0.01 & 0.01 & 0.01 \\
& Attack simulation for constraint generation & 4813.79 & 3740.89 & 5195.99 & 3807.08 \\
\midrule
\multicolumn{2}{c||}{Watermark-database generation (executed once; results reusable across datasets)} & \multicolumn{4}{c}{51.90} \\
\midrule
\multirow{2}{*}{\makecell{Watermarked-table\\synthesis (online)}} & Watermarked-histogram generation & 156.94 & 155.47 & 155.29 & 156.33 \\
& Table synthesis & 12.31 & 4.02 & 4.28 & 4.20 \\
\midrule
\multirow{2}{*}{\makecell{Watermark decoding\\(online)}} & Watermarked-histogram extraction & 5.79 & 2.09 & 4.45 & 3.34 \\
& Partial-order extraction and data buyer identification & 0.01 & 0.01 & 0.01 & 0.01 \\
\bottomrule
\end{tabular}
}
\end{table*}

\subsubsection{Analysis on $N$}

We made the following observations from Figure~\ref{fig:experiments:num_users}. Firstly, traceability accuracy decreased as $N$ increased. This was because the bit error threshold $\delta_{BE}$ decreased so that the $L$-bit watermark space could accommodate more data buyers (see Equation~\ref{eq:FPR2}), resulting in a reduced bit error tolerance. Secondly, the increase in $N$ had only a minimal impact on RAQE. Based on these observations, we recommend that the data owner set $N$ to a small number as long as $N$ is greater than the number of potential data buyers.

\subsubsection{Analysis on $L$}

We made the following observations from Figure~\ref{fig:experiments:num_bits}. Firstly, traceability accuracy increased slightly with $L$. This was because $\delta_{BE}$ increased due to the enlarged watermark space (see Equation~\ref{eq:FPR2}), leading to an improved bit error tolerance. Secondly, changes in $L$ had a minimal impact on RAQE. Note that a 32-bit watermarking scheme can already support a large number of data buyers (when $N=10^5$, the traceability accuracy was almost 1 under 5\% adaptive tuple-deletion attacks while the utility was close to the non-watermarked synthetic table; see Figure~\ref{fig:experiments:num_users}), we recommend that the data owner set $L$ to 32 as in our default configuration and balance robustness and utility by configuring robustness parameters as discussed in Section~\ref{sec:experiments:constraint}.

\subsection{Evaluation on Runtime Efficiency} \label{sec:experiments:time}

To answer \textbf{RQ5}, we measured the runtime of each stage in \oursys and the results are shown in Table~\ref{tab:experiments:time}. It can be observed that the online overhead of \oursys is modest and practical for real-world deployment.

\section{Related Work}

\noindent \textbf{Cell-Based Watermarking Schemes.} Some schemes encoded watermarks into the least significant bits (LSB) of cells~\cite{LSB1,LSBSYS,LSBMulti,LSBQuality,LSBPrivacy}. However, these schemes were limited in robustness because LSBs were sensitive to small perturbations. Therefore, another line of work~\cite{PKMeanStdNoise,LSBRobust,TabularMark} encoded watermarks into specific noise patterns and injected the noise into certain cells. \myhi{However, these schemes were limited in utility because injecting noise into cells could compromise tuple integrity. In contrast, \oursys preserves tuple integrity and optimizes tuple-frequency distributions, thus having better utility.}

\noindent \textbf{Statistic-Based Watermarking Schemes.} They encoded watermarks into table statistics, such as column densities~\cite{MarkerTuple}, value frequencies~\cite{Categorical1,PKReversibleCircularHistogram,PKFPM1HistogramShifting}, tuple frequencies~\cite{FreqWM}, and values of specific functions~\cite{PKMSBMinMaxMeanStdHidingFunction}. To balance robustness and utility, these schemes imposed constraints on table modifications during watermark encoding. Specifically, some schemes imposed utility constraints, for example, on attribute means~\cite{PKMSBMinMaxMeanStdHidingFunction}, value ranges~\cite{PKFPM1HistogramShifting}, results of predefined classification tasks~\cite{MIModel}, and partition histograms~\cite{FreqWM}. \myhi{However, constraints imposed by these schemes~\cite{PKMSBMinMaxMeanStdHidingFunction,PKFPM1HistogramShifting,MIModel} were limited to specific applications, which was insufficient to support diverse machine learning and statistical analysis tasks for which synthetic tables were used, or were too restrictive (e.g., the ranking constraint in~\cite{FreqWM}), which weakened traceability and robustness.} For example, the traceability accuracy of FreqyWM~\cite{FreqWM} was almost zero under attacks. Moreover, some work~\cite{PKReversibleCircularHistogram} imposed robustness constraints, but \myhi{considered a rather limited set of attacks (e.g., only deletion and insertion attacks in~\cite{PKReversibleCircularHistogram}), which undermined robustness.} \myhi{In contrast,\! \oursys designs partial-order patterns within the histogram channel and proposes the constraint generation mechanism, offering the data owner with an effective robustness-utility trade-off in the context of various attacks.}

\noindent \textbf{Generative Watermarking Schemes.} They encoded watermarks into pseudo-random sequences used for table synthesis. For example, TabWak~\cite{TabWak} encoded the watermark into Gaussian noise vectors and input these vectors into DDIM~\cite{DDIM} (an invertible diffusion model) to synthesize the watermarked table, while RINTAW~\cite{RINTAW} and MUSE~\cite{MUSE} used a pseudo-random function to associate tuples with scores, encoded the watermark into the sum of tuple scores, and biased the table synthesis process toward tuples with specific scores. \myhi{However, these schemes were limited in traceability.} For example, the traceability accuracy of TabWak was nearly zero when encoding 32-bit watermarks to track 1000 data buyers. This was because pseudo-random sequences were difficult to recover accurately to support a large number of watermark bits. \myhi{Different from these schemes, \oursys designs partial-order patterns that can be accurately decoded for multi-bit watermarks.}

\section{Conclusion} \label{sec:conclusion}

We have studied the problem of effectively tracing data buyers for selling synthetic tabular data in the data market. To this end, we have proposed \oursys, a novel watermarking scheme that encodes multi-bit watermarks into partial-order patterns within the cluster-histogram channel. To provide the data owner with an effective robustness-utility trade-off, we have proposed the constraint generation mechanism that formulates the watermark encoding process as a constrained optimization problem, to provide theoretical guarantees for the data owner's robustness requirement. We have designed effective optimization mechanisms to solve this problem to synthesize high-utility watermarked tables. We have conducted experiments on four widely used real-world datasets, and the results have shown that \oursys significantly outperforms the state-of-the-art baselines. Moreover, we have conducted extensive parameter analysis and derived practical configuration guidelines. Finally, we have evaluated the runtime efficiency of \oursys and demonstrated its practicality for real-world deployment.

In future work, we aim to extend \oursys to support watermarking for multi-table databases and to address collusion attacks, where coordinated attackers combine their watermarked copies to detect and remove watermark patterns.

\bibliographystyle{spmpsci}
\bibliography{sample}

\end{document}